\documentclass[structabstract]{aa} 
\usepackage{graphicx}
\usepackage{txfonts}
\usepackage{natbib}
\usepackage{color}
\defcitealias{mk98}{MK98}
\begin{document}
\title{Quadruple-peaked spectral line profiles as a tool to constrain gravitational potential of shell galaxies}

\author{I. Ebrov\'{a}\inst{1}\fnmsep\inst{2}
\and
L. J\'{i}lkov\'{a}\inst{3}\fnmsep\inst{4}
\and
B. Jungwiert\inst{1}
\and
M. K\v{r}\'{i}\v{z}ek\inst{1}\fnmsep\inst{5}
\and 
M. B\'{i}lek\inst{1}\fnmsep\inst{2}
\and
K. Barto\v{s}kov\'{a}\inst{1}\fnmsep\inst{3}
\and
T. Skalick\'{a}\inst{3}
\and
I.~Stoklasov\'{a}\inst{1}
}

\institute{Astronomical Institute, Academy of Sciences of the Czech Republic, Bo\v{c}n\'{i} II 1401/1a, CZ-141 31 Prague, Czech Republic\\
\email{ivana@ig.cas.cz}
\and
Faculty of Mathematics and Physics, Charles University in Prague, Ke~Karlovu 3, CZ-121 16 Prague, Czech Republic
\and 
Department of Theoretical Physics and Astrophysics, Faculty of Science, Masaryk University, Kotl\'a\v{r}sk\'a 2, CZ-611\,37 Brno, Czech Republic
\and
ESO, Alonso de Cordova 3107, Casilla 19001, Santiago, Chile
\and
Astronomical Institute, Faculty of Mathematics and Physics, Charles University in Prague, V Hole\v{s}ovi\v{c}k\'{a}ch 2, CZ-180 00 Prague, Czech Republic
}

\date{Received 2 June 2012; accepted 9 June 2012}

\abstract
{Stellar shells observed in many giant elliptical and lenticular as well as a few spiral and dwarf galaxies presumably result from galaxy mergers. Line-of-sight velocity distributions of the shells could, in principle, if measured with a sufficiently high signal-to-noise ratio, constitute a method to constrain the gravitational potential of the host galaxy.}
{Merrifield \& Kuijken (1998) predicted a double-peaked line profile for stationary shells resulting from a nearly radial minor merger. In this paper, we aim at extending their analysis to a more realistic case of expanding shells, inherent to the merging process, whereas we assume the same type of merger and the same orbital geometry.}
{We used an analytical approach as well as test particle simulations to predict the line-of-sight velocity profile across the shell structure. Simulated line profiles were convolved with spectral PSFs to estimate peak detectability.}
{The resulting line-of-sight velocity distributions are more complex than previously predicted due to nonzero phase velocity of the shells. In principle, each of the Merrifield \& Kuijken (1998) peaks splits into two, giving a quadruple-peaked line profile, which allows more precise determination of the potential of the host galaxy and contains additional information. We find simple analytical expressions that connect the positions of the four peaks of the line profile and the mass distribution of the galaxy, namely, the circular velocity at the given shell radius and the propagation velocity of the shell. The analytical expressions were applied to a test-particle simulation of a radial minor merger, and the potential of the simulated host galaxy was successfully recovered. Shell kinematics can thus become an independent tool to determine the content and distribution of the dark matter in shell galaxies up to $\sim$100\,kpc from the center of the host galaxy.}
{}


\maketitle


\section{Introduction}

Several methods have been used to measure the gravitational potentials and their gradients of elliptical galaxies, including strong and weak gravitational lensing \citep[e.g.,][]{gavazzi07,mandelbaum08,auger10}, X-ray observations of hot gas in the massive gas-rich galaxies \citep[e.g.,][]{fukazawa06,nagino09,churazov08,das10}, rotation curves of detected disks and rings of neutral hydrogen \citep[e.g.,][]{weijmans08}, stellar-dynamical modeling from integrated light spectra \citep[e.g.,][]{weijmans09,delorenzi09,churazov10,thomas11}, and the use of tracers such as planetary nebulae, globular clusters, and satellite galaxies \citep[e.g.,][]{coccato09,nierenberg11,deason12,norris12}. All these methods have their limits, such as the redshift of the observed object, the luminosity profile, and gas content. In particular, the use of stellar dynamical modeling is plausible in the wide range of galactic masses, as long as spectroscopic data are available. However, it becomes more challenging beyond a few optical half-light radii. Other complementary gravitational tracers or techniques are required to derive mass profiles in outer parts of the galaxies. When comparing independent techniques for the same objects at similar galactocentric radii, discrepancies in the estimated circular velocity curves were revealed together with several interpretations \citep[e.g.,][]{churazov10,das10}. The compared techniques usually employ modeling of the X-ray emission of the hot gas (assuming hydrostatic equilibrium) and dynamical modeling of the optical data in the massive early-type galaxies. Therefore, even for the most massive galaxies with X-ray observations available, there is a need for other methods to independently constrain the gravitational potential at various radii.

Shell galaxies are galaxies that contain arc-like fine features, which were first noticed by \citet{arp66}. These structures are made of stars and form open, almost concentric arcs that do not cross each other. Shells are relatively common in elliptical or lenticular galaxies. At least 10\,\% of all these galaxies in the local universe possess shells. Nevertheless, shells occur markedly most often in regions of low galaxy density, and perhaps up to half of E and S0 galaxies in these environments are shell galaxies \citep{malin83,schweizer83,schweizer85,colbert01}. Shells can also be associated with dust \citep{sikkema07,stickel04} and neutral hydrogen emission \citep{schiminovich94,schiminovich95,balcells96,petric97,horellou01}. In addition, \citet{charmandaris00} detected the presence of dense molecular gas in the shells of NGC 5128.

Shells are thought to be by-products of minor mergers of galaxies \citep{quinn84}, although they can also be formed during major mergers \citep{hernquist92}. The most regular shell systems are believed to result from nearly radial mergers \citep{dupraz86,hernquist88}. When a small galaxy enters the sphere of influence of a big elliptical galaxy on a radial or close-to-radial trajectory, it disintegrates and its stars begin to oscillate in the potential of the big galaxy. At their turning points, the stars have the lowest speed and thus tend to spend most of the time there, where they pile up and produce arc-like structures in the luminosity profile of the host galaxy when viewed perpendicular to the axis of the collision.

Measurement of the number and distribution of shells can, in principle, yield to an approximate estimate of the mass distribution of the host galaxy and the time since the merger \citep{quinn84,dupraz86,hernquist87a,hernquist87b,canalizo07}. But both of these observables are sensitive to details such as the dynamical friction and the gradual decay of the cannibalized galaxy during the merger \citep{dupraz87,james87,heisler90,ebrova10}. 
Moreover, if the core of the cannibalized galaxy survives the merger, new generations of shells are added during each successive passage. This was predicted by \citet{dupraz87} and successfully reproduced by \citet{katka11} in self-consistent simulations. All these effects complicate the simulations to such an extent that the interest in shell galaxies largely faded by the end of the 1980s. Recently, this topic has raised interest again, thanks to the discovery of shells in a quasar host galaxy \citep{canalizo07} and shell structures in M31 \citep{fardal07,fardal08} and in the Fornax dwarf \citep{coleman04}. \citet{helmi03} suggested that ring-like stellar structures, including the one observed in the outer disk of the Milky Way (the so-called Monoceros ring), could be analogous to shells. A significant number of shells is also contained in the early-type galaxy sample of the ongoing ATLAS$^{\mathrm{3D}}$ project, including images of galaxies with a surface brightness down to 29\,mag$/$arcsec$^2$ \citep[see, e.g.,][]{krajnovic11,duc11}. \citet{kim12} identified shells in about 6\% of a sample of 65 early-type galaxies from the Spitzer Survey of Stellar Structure in Galaxies (S$^4$G). Shells also appear to be suitable for indirect detection of dark matter via gamma-ray emission from dark matter self-annihilations \citep{sanderson12}. About 70\,\% of a complete sample of nearby (15--50\,Mpc) luminous ($M_{\mathrm{B}}<-20$\,mag) elliptical galaxies were found to show tidal features by Tal et al. (2009). Faint structures, including shells and other signatures of recent gravitational interaction (tidal tails and streams), were found in the Sloan Digital Sky Survey (SDSS). \citet{kaviraj10} identified 18\,\% of early-type galaxies (ETGs) in the SDSS Stripe82 sample as having disturbed morphologies; similarly, \citet{miskolczi11} found tidal features in 19\,\% of their sample of galaxies from SDSS DR7. Observations of warm gas by \citet{rampazzo03} in five shell galaxies showed irregular gaseous velocity fields (e.g., a double nucleus or elongated gas distribution with asymmetric structure relative to the stellar body), and in most cases, gas and stellar kinematics appear decoupled. \citet{rampazzo07}, \citet{marino09}, and \citet{trinchieri08} investigated star formation histories and hot gas content using the NUV and FUV Galaxy Evolution Explorer (GALEX) observations (and in the latter case also X-ray ones) in a few shell galaxies. The results support accretion events in the history of shell galaxies.

\citet{mk98}, hereafter \citetalias{mk98}, studied theoretically the kinematics of a stationary shell, a monoenergetic spherically symmetric system of stars oscillating on radial orbits in a spherically symmetric potential. They predicted that spectral line profiles of such a system exhibit two clear maxima, which provide a direct measure of the gradient of the gravitational potential at the shell radius. The first attempt to analyze the kinematical imprint of a shell observationally was made by \citet{romanowsky12}, who used globular clusters as shell tracers in the early-type galaxy M87. \citet{fardal12} obtained radial velocities of giant stars in the so-called western shelf in M31 Andromeda galaxy. They successfully analyzed the shell pattern in the space of velocity versus radius. 

Nevertheless, real-world shells are not stationary features. The stars of the satellite galaxy have a continuous energy distribution, and so at different times, the shell edge is made of stars of different energies, as they continue to arrive at their respective turning points. Thus, the shell front appears to be traveling outwards from the center of the host galaxy and shell spectral-line profiles are more complex (\citealt{jilkova10}, \citealt{ebrova11}, see also \citealt{fardal12}).

In this paper, we derive spectral-line profiles of nonstationary shells. We assume that shells originate from radial minor mergers of galaxies, as proposed by \citet{quinn84}. We find that both of the original \citetalias[][]{mk98} peaks in the spectral line are split into two, resulting in a quadruple shape, which can still be used to constrain the host galaxy potential and even bring additional information. We outline the simplified theoretical model and derive the shell velocities in Sect.~\ref{sub:rad_osc}, and describe the origin of the quadruple line profile in Sect.~\ref{sec:4peak}. In Sect.~\ref{sec:app}, we derive equations connecting the observable features of the quadruple-peaked line-of-sight velocity distribution (LOSVD) with parameters of the host galaxy potential in the vicinity of the shell edge. We compare these analytical predictions with the theoretical model (Sect.~\ref{sec:Compars}) and with results of test-particle simulations of the radial minor merger (Sect.~\ref{sec:N-Simulations}). Section~\ref{sec:N-Simulations} also demonstrates the derivation of the galactic potential from the simulated spectral data.


\section{Model of radial oscillations} \label{sub:rad_osc}

If we approximate the shell system with a simplified model, we can describe its evolution completely depending only on the potential of the host galaxy. The approximation lies in the numerical integration of radial trajectories of stars in a spherically symmetric potential. The distribution of energies of stars is continuous, and these stars were released from a small volume in the phase space. We call it the model of radial oscillations, and it corresponds to the notion that the cannibalized galaxy came along a radial path and disintegrated in the center of the host galaxy. As a result the stars were released at one moment in the center and began to oscillate freely on radial orbits. This approach was first used by \citet{quinn84}, followed by \citet{dupraz86,dupraz87} and \citet{hernquist87a,hernquist87b}.

\subsection{Turning point positions and their velocities} \label{sub:TP}

In shell galaxies, the shells are traditionally numbered according to the serial number of the shell, $n$, from the outermost to the innermost (which in the model of radial oscillations for a single-generation shell system corresponds to the oldest and the youngest shell, respectively). If the cannibalized galaxy comes from the right side of the host galaxy, stars are released in the center of the host galaxy. After that, they reach their apocenters for the first time. But a shell does not form here yet, because the stars are not sufficiently phase wrapped. We call this the zeroth oscillation (the zeroth turning point) as we try to match the number of oscillations with the customary numbering scheme of the shells. We label the first shell that occurs on the right side (the same side from which the cannibalized galaxy approached) with $n=1$. Shell no.~2 appears on the left side of the host galaxy, no.~3 on the right, and so forth. 

In the model of radial oscillations, the shells occur close to the radii where the stars are located in their apocenters at a given moment (the current turning point, $r_{\mathrm{TP}}$, in our notation). The shell number $n$ corresponds to the number of oscillations that the stars near the shell have completed or are about to complete. The current turning point $r_{\mathrm{TP}}$ must follow the equation
\begin{equation}
t=(n+1/2)T(r_{\mathrm{TP}}),
\label{eq:Trn}
\end{equation}
where $t$ is the time elapsed since stars were released in the center of the host galaxy. $T(r)$ is the period of radial motion at a galactocentric radius $r$ in the host galaxy potential $\phi(r)$:
\begin{equation}
T(r)=\sqrt{2}\int_{0}^{r}\left[\phi(r)-\phi(r')\right]^{-1/2}\mathrm{d}r'.
\label{eq:Tr}
\end{equation}

The position of the current turning point evolves in time with a velocity given by the derivative of Eq.~(\ref{eq:Trn}) with respect to radius
\begin{equation}
v_{\mathrm{TP}}(r;n)=\mathrm{\mathrm{d}r/d}t=\frac{1}{\mathrm{\mathrm{d}t/d}r}=\frac{1}{n+1/2}\left(\mathrm{d}T(r)/\mathrm{d}r\right)^{-1}.
\label{eq:vTP}
\end{equation}
We can clearly see from this relation, which was first derived by \citet{quinn84}, that any further turning point (turning point with higher $n$) at the same radius moves more slowly than the former one. Thus causes a gradual densification of the space distribution of the shell system with time.

Technically, the reason for this densification is that the time difference between the moments when two stars with similar energy reach their turning points is cumulative. Let $\bigtriangleup t$ be the difference in periods at two different radii $r_{\mathrm{a}}$ and $r_{\mathrm{b}}$ (with $r_{\mathrm{a}}<r_{\mathrm{b}}$, on the right). The radius where stars complete the first oscillation moves from $r_{\mathrm{a}}$ to $r_{\mathrm{b}}$ in $\bigtriangleup t$. But in the second orbit on the left, the stars from $r_{\mathrm{b}}$ will already have a lag of $\bigtriangleup t$ behind those from $r_{\mathrm{a}}$ and will just be getting a second one, so the third one (the second on the same side) reaches $r_{\mathrm{b}}$ from $r_{\mathrm{a}}$ in $2\times\bigtriangleup t$. Every $n$th completed oscillation on the right side, then moves $n$ times more slowly than the first one. The situation is similar on the left side, and the shell system is getting denser. Moreover, the turning point has an additional lag of $1/2T(r_{\mathrm{TP}})$, because the stars were released in the center of the host galaxy before their zeroth oscillation. This is the source of the factor $(n+1/2)$ in Eqs.~(\ref{eq:Trn}) and (\ref{eq:Tr}).

\begin{figure*}
\centering
\includegraphics[width=6.5cm]{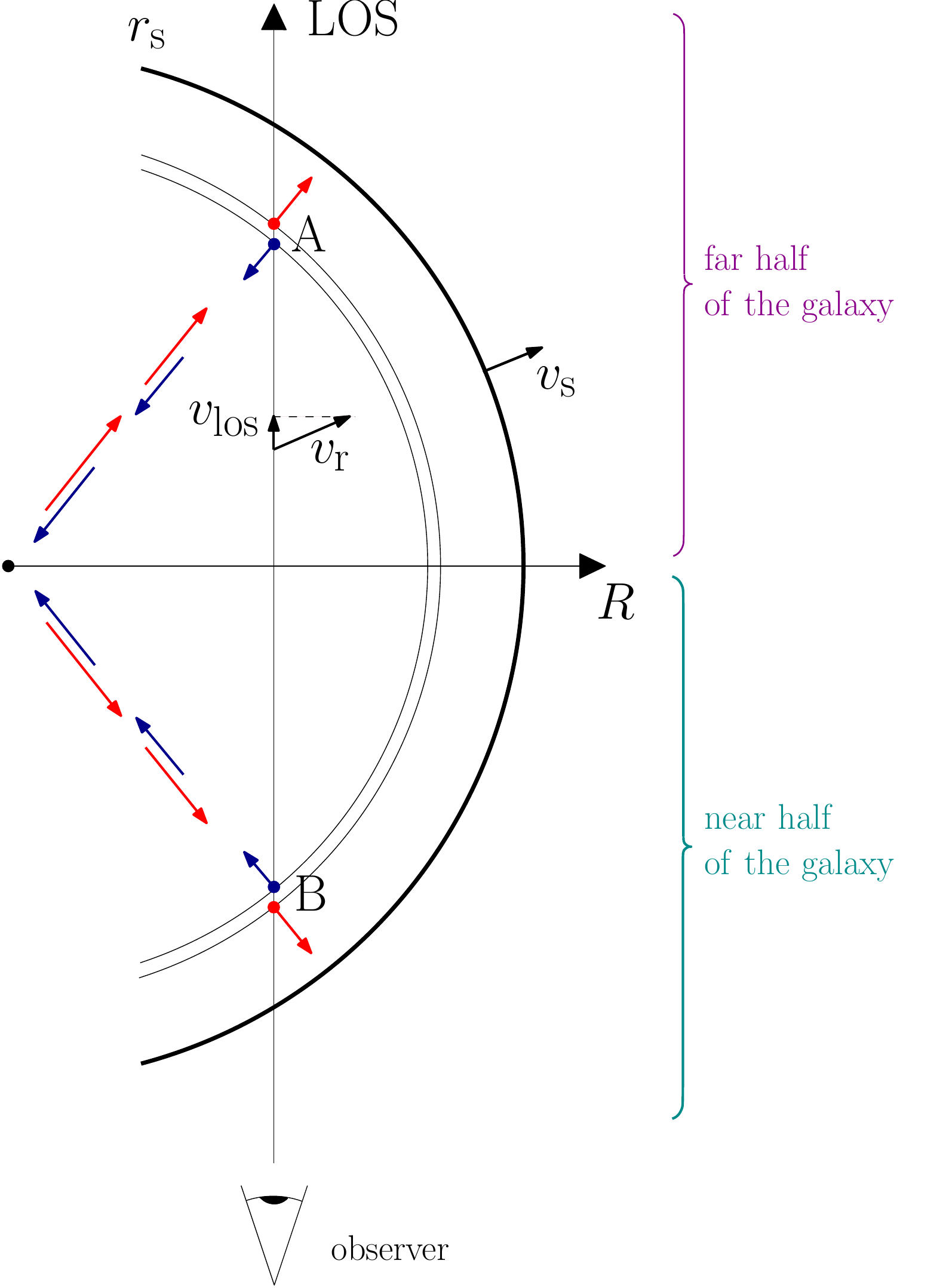}
\includegraphics[width=6.0cm]{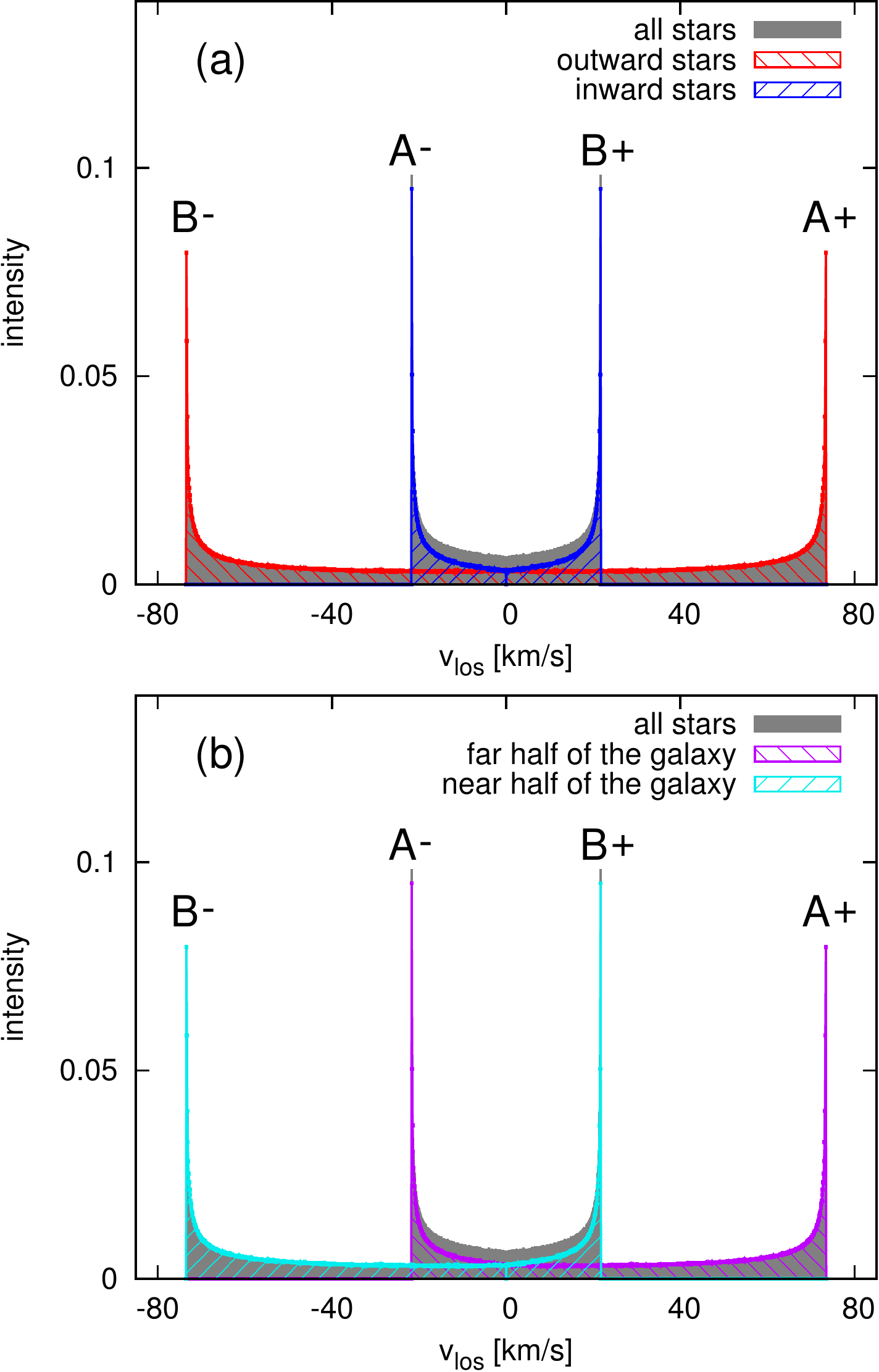}
\caption{Kinematics of a moving shell. Compare with Fig.~3 in \citetalias{mk98} for a stationary shell. Left: Scheme of the kinematics of a shell with radius $r_{\mathrm{s}}$ and phase velocity $v_{\mathrm{s}}$. The shell is composed of stars on radial orbits with radial velocity $v_{\mathrm{r}}$ and LOS velocity $v_{\mathrm{los}}$. Right: The LOSVD at projected radius $R=0.9r_{\mathrm{s}}$, where $r_{\mathrm{s}}=120$\,kpc (parameters of the shell are highlighted in bold in Table~\ref{tab:param}), in the framework of the model of radial oscillations (Sect.~\ref{sub:LOSVD-rad}). The profile does not include stars of the host galaxy, which are not part of the shell system, and is normalized, so that the total flux equals one. (a) The LOSVD showing separate contributions from inward and outward stars; (b) the same profile, separated for contributions from the half of the host galaxy closer to the observer (the one including point B) and the more distant half (includes point A).}
\label{fig:Mr.Eggy}\label{fig:anath}
\end{figure*}


\subsection{Real shell positions and velocities} \label{sub:edge}
Even in the framework of the radial oscillation model, the position and velocity of the true edge of the shell cannot be expressed in a straightforward manner. Photometrically, shells appear as a brightening in the luminosity profile of the galaxy with a sharp cut-off. This is because the stars of the cannibalized galaxy occupy a limited volume in the phase space. With time, the shape of this volume gets thinner, more elongated, and wrapped around invariant surfaces defined by the trajectories of the particles, increasing its coincidence with these surfaces. A shell appears close to the points where the invariant surface is perpendicular to the plane of the sky \citep{nulsen89}. For the $n$th shell, this is the largest radius where stars about to complete their $n$th oscillation are currently located. This radius is always larger than that of the current turning point of the stars that are completing their $n$th oscillation. Thus, the shell edge consists of outward-moving stars about to complete their $n$th oscillation.

\citet{dupraz86} state that the stars forming the shell move with the phase velocity of the shell. While we show that this holds only approximately, we use this equality in Sect.~\ref{sec:app} to derive the relation between the shell kinematics and the potential of the host galaxy.

The position of a star, $r_{\mathrm{*}}$, at a given time $t$ since the release of the star in the center of the host galaxy is given by an implicit equation for $r_{\mathrm{*}}$ and is a function of the star energy, or equivalently the position of its apocenter $r_{\mathrm{ac}}$.  \footnote{We denote the apocenter of the star corresponding to its energy as $r_{\mathrm{ac}}$, whereas $r_{\mathrm{TP}}$ (the current turning point) is the radius at which the stars reach their apocenters at the time of measurement.}
For stars with the integer part of $t/[2T(r_{\mathrm{ac}})]$ odd, the equation reads:
\begin{equation}
\begin{array}{rcl}
t=(n+1)\sqrt{2} & \int_{0}^{r_{\mathrm{ac}}} & \left[\phi(r_{\mathrm{ac}})-\phi(r')\right]^{-1/2}\mathrm{d}r'- \\
- & \int_{0}^{r_{\mathrm{*}}} & \left[2(\phi(r_{\mathrm{ac}})-\phi(r'))\right]^{-1/2}\mathrm{d}r'.
\end{array}
\label{eq:r*-}
\end{equation}
For stars that have completed an even number of half-periods (only such stars are found on the shell edge), the equation is
\begin{equation}
\begin{array}{rcl}
t=n\sqrt{2} & \int_{0}^{r_{\mathrm{ac}}} & \left[\phi(r_{\mathrm{ac}})-\phi(r')\right]^{-1/2}\mathrm{d}r'+ \\
+ & \int_{0}^{r_{\mathrm{*}}} & \left[2(\phi(r_{\mathrm{ac}})-\phi(r'))\right]^{-1/2}\mathrm{d}r'.
\end{array}
\label{eq:r*+}
\end{equation}
The first term in Eq.~(\ref{eq:r*+}) corresponds to $n$ radial periods for the star's energy ($n$ is maximal so that $nT(r_{\mathrm{ac}})<t$), while the other term corresponds to the time that it takes to reach radius $r_{\mathrm{*}}$ from the center of the galaxy. Even for the simplest galactic potentials, these equations are not analytically solvable and must be solved numerically. 

The position of the $n$th shell $r_{\mathrm{s}}$ equals the maximal radius $r_{\mathrm{*,max}}$ that solves Eq.~(\ref{eq:r*+}) for the given $n$. The shell velocity $v_{\mathrm{s}}$ is obtained from the numerical derivative of a set of values of $r_{\mathrm{*,max}}$ for several close values of $t$. 

The stellar velocity at the shell edge is obtained by inserting $r_{\mathrm{*,max}}$ with its corresponding $r_{\mathrm{ac}}$ into:
\begin{equation}
v(r_{\mathrm{*}})=\pm\sqrt{2[\phi(r_{\mathrm{ac}})-\phi(r_{\mathrm{*}})]}.
\label{eq:v*}
\end{equation}
For the stars following Eq.~(\ref{eq:r*+}), the velocity will be positive; for the rest, it will be negative.

It is clear that $v(r_{\mathrm{*,max}})\leq v_{\mathrm{s}}$. Actually, $v(r_{\mathrm{*,max}})$ is always slightly lower than the phase velocity of the shell (Table~\ref{tab:param}). Meanwhile, the position of the shell for a given time is not far from the current turning point, and their separation changes slowly. Thus, the velocity of the turning points given in Eq.~(\ref{eq:vTP}) is a good approximation for the shell velocity (Fig.~\ref{fig:vs}). Equation~(\ref{eq:vTP}) is not generally solvable analytically either, but the numerical calculation of $v_{\mathrm{TP}}$ is much easier than determining the true velocity $v_{\mathrm{s}}$ as described in this chapter.


\subsection{Kinematics of shell stars}\label{sub:LOSVD-rad}

In the same model, we can also describe the LOSVD of a shell at a given time $t$, for a given potential of the host galaxy $\phi(r)$. In this paper, we model the host galaxy potential of the host galaxy as a double Plummer sphere, as described in Sect.~\ref{sec:param}.

Eqs.~(\ref{eq:r*-}) and (\ref{eq:r*+}) give the actual star position $r_{\mathrm{*}}$ and the shell number $n$ for any apocenter $r_{\mathrm{ac}}$ in a range of energies. The radial velocity of a star on the particular radius is given by inserting the corresponding pair of $r_{\mathrm{ac}}$ a $r_{\mathrm{*}}$ in Eq.~(\ref{eq:v*}). Naturally, the projections of these velocities to the selected line-of-sight (LOS) form the LOSVD.
To reconstruct the LOSVD, we have to add an assumption about the behavior of shell brightness in time. In Sect.~\ref{sub:LOSVD-app}, we show our choice of the behavior and also illustrate that the particular choice does not matter much.


\section{Quadruple-peaked LOSVD}\label{sec:4peak}

Fig.~\ref{fig:Mr.Eggy} illustrates a measurement of the LOSVD of stars in the shell, which is  composed of inward and outward stars on radial trajectories. The stars near the edge of the shell move slowly. But it is clear from the geometry that contributions add up from different galactocentric distances, where the stars are either still traveling outwards to reach the shell or returning from their apocenters to form a nontrivial LOSVD. \citetalias{mk98} showed that the maximal contribution to the LOSVD comes from stars at two particular locations along the line of sight (A and B), both of which are at the same galactocentric distance. 

In \citetalias[][]{mk98}'s stationary shell model, inward stars at the same radius differ from outward stars only in the sign of the LOS velocity $v_{\mathrm{los}}$. This is not true when the edge of the shell moves outwards with velocity $v_{\mathrm{\mathrm{s}}}$. At any given instant, the stars that move inwards are returning from a point where the shell edge was at some earlier time, and so their apocenter is inside the current shell radius $r_{\mathrm{s}}$. Similarly, the stars that move outwards will reach the shell edge in the future. Consequently, the stars that move inwards are always closer to their apocenter than those moving outwards at the same radius, and their velocity is thus smaller. The inward stars move toward the observers in the farther of the two \citetalias[][]{mk98} points (A) and away from them in the nearer point (B), while the stars moving outwards behave in the opposite manner. Together, there are four possible velocities with the maximal contribution to the LOSVD, resulting in its symmetrical quadruple shape shown in Fig.~\ref{fig:anath}. In fact, for a moving shell, points A and B are not at the same galactocentric radius for inward and outward stars. For inward stars, points A and B are a little closer to the center as indicated in Fig.~\ref{fig:Mr.Eggy}. This is discussed in Sect.~\ref{sub:rvmax}.

In the right-hand panel of Fig.~\ref{fig:anath}, we used the model of radial oscillations as described in Sect.~\ref{sub:LOSVD-rad} to illustrate individual contributions to the LOSVD. \citetalias{mk98} constructed an analytical function describing the LOSVD close to the edge of their stationary shell model. This function exhibits intensity maxima that coincide with maximal/minimal velocity, leading to the symmetrical double-peaked profile with peaks at the edges of the LOSVD. This cannot be shown for a general moving shell, but Fig.~\ref{fig:anath} demonstrates that the intensity maxima coincide with velocity extremes for separate contributions to the LOSVD.

\begin{figure}
\centering{}
\resizebox{\hsize}{!}{\includegraphics{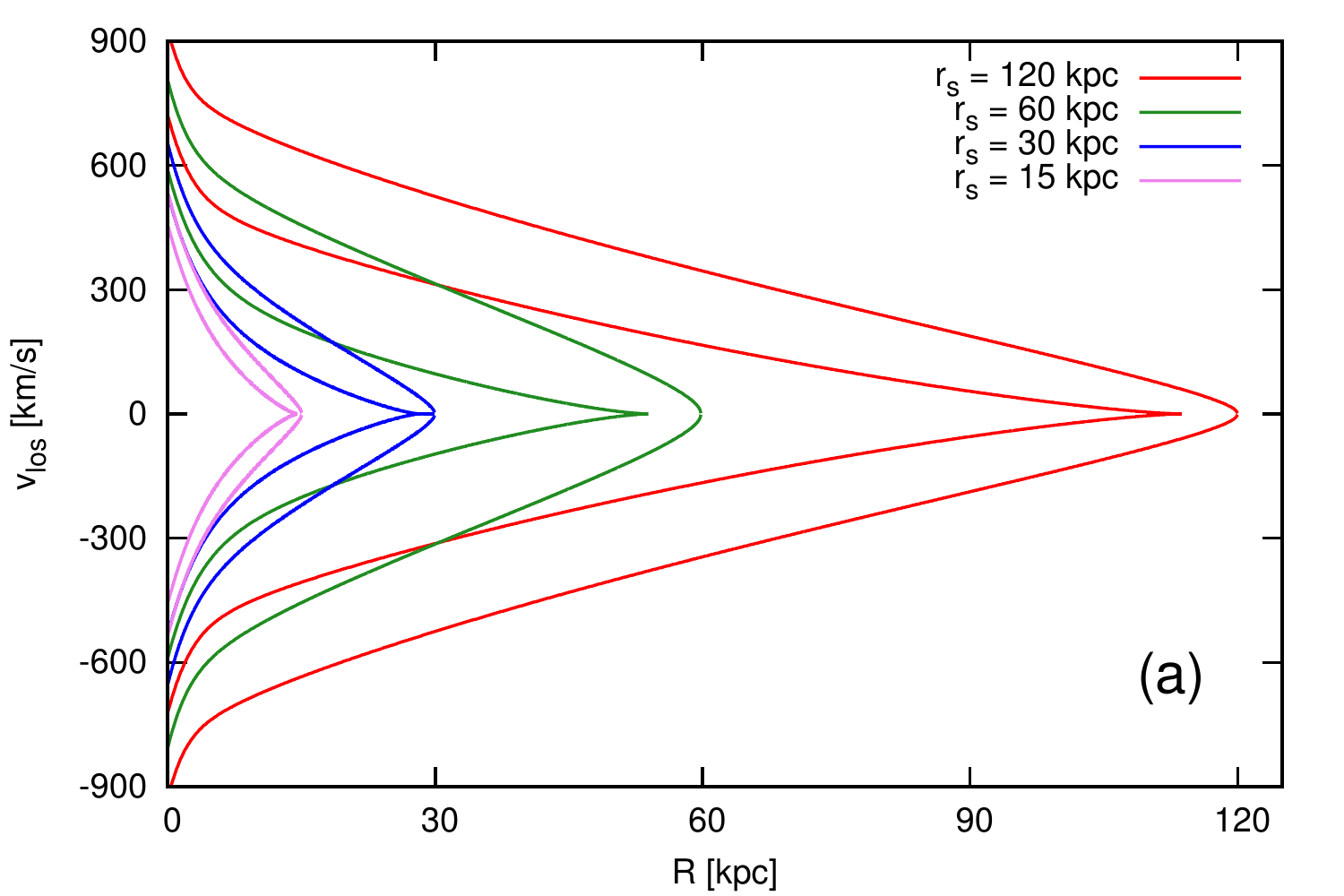}}\\
\resizebox{\hsize}{!}{\includegraphics{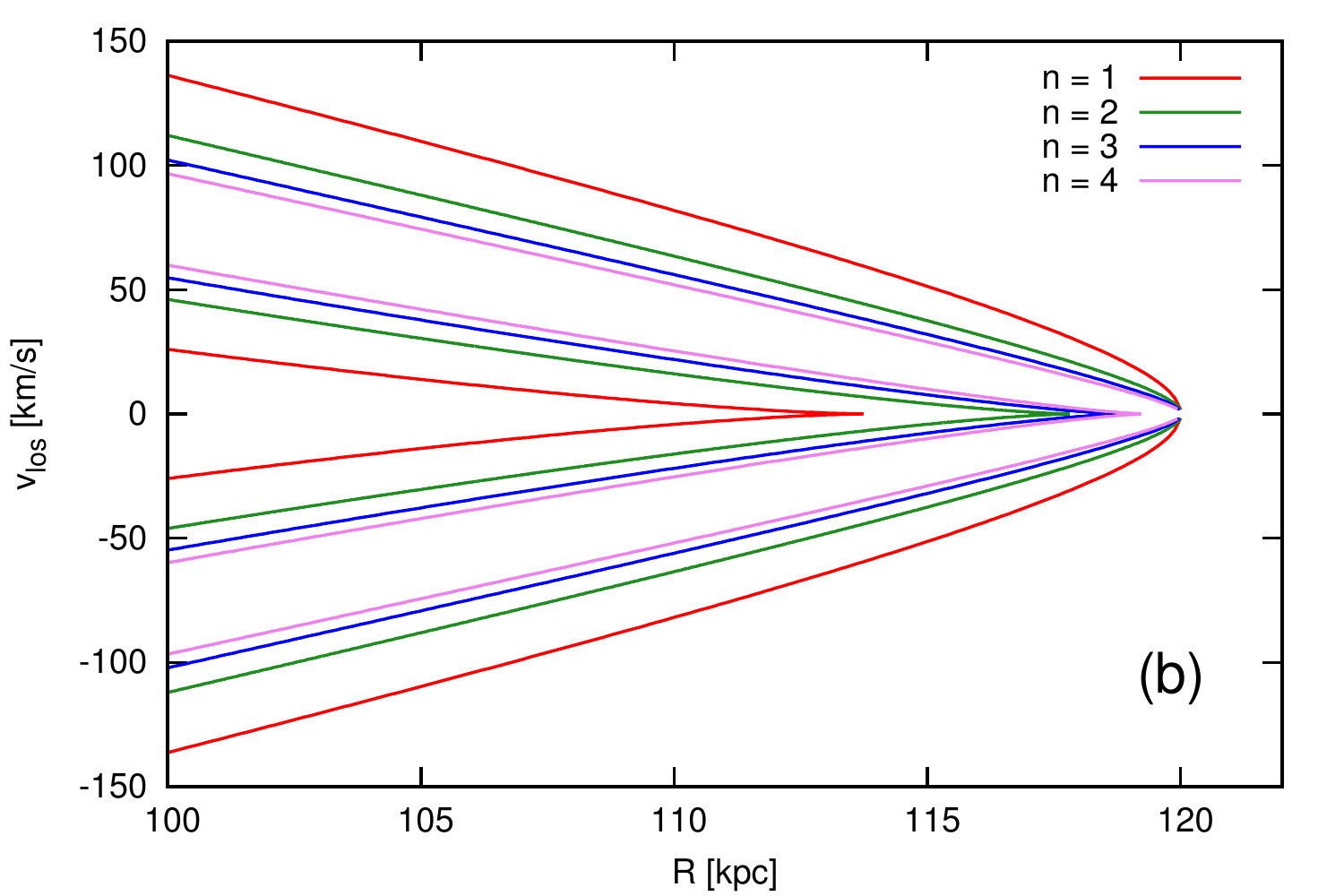}}
\caption{Locations of peaks of the LOSVDs in the framework of the model of radial oscillations (Sect.~\ref{sub:LOSVD-rad}): (a) for the first shell at different radii, (b) for the first to the fourth shell at the radius of 120\,kpc. Parameters of all shells are shown in Table~\ref{tab:param}. For parameters of the host galaxy potential, see Sect.~\ref{sec:param}.}
\label{fig:radmax}
\end{figure}

The separation in velocity between peaks for a given projected radius $R$ is given by the distance of $R$ from the edge of the shell $r_{\mathrm{s}}$. The profile shown in Fig.~\ref{fig:anath} corresponds to projected radius $R=0.9r_{\mathrm{s}}$. The closer to the shell edge, the narrower the profile is. The separation of the peaks at a given $R$ depends on the phase velocity of the specific shell, near which we observe the LOSVD. This velocity is, for a fixed potential, given by the shell radius and its serial number (Sect.~\ref{sub:TP}). These effects are illustrated in Fig.~\ref{fig:radmax}, where we show the positions of the LOSVD peaks for the first shell at different radii $r_{\mathrm{s}}$ and for a shell at 120\,kpc with different serial numbers $n$. Note that the higher the serial number $n$ at a given radius, the smaller is the difference in the phase velocity between the two shells with consecutive serial numbers and thus in the positions of the respective peaks. Parameters of the corresponding shells can be found in Table~\ref{tab:param}. 

\begin{table}
\centering
\caption{Parameters of shells for which the LOSVD intensity maxima are shown in Fig.~\ref{fig:radmax}.}
\label{tab:param}
\begin{tabular}{cccccccc}
\hline\hline
$t$ & $n$ & $r_{\mathrm{s}}$ & $r_{\mathrm{TP}}$ & $v_{\mathrm{s}}$ & $v(r_{\mathrm{*,max}})$ & $v_{\mathrm{TP}}$ & $v_{\mathrm{c}}$ \\
Myr &   & kpc & kpc & km$/$s & km$/$s & km$/$s & km$/$s \\
\hline 
215 & 1 & 15 & 14.5 & 63.5 & 57.5 & 61.2 & 245\\
416 & 1 & 30 & 28.3 & 90.3 & 82.6 & 81.0 & 261\\
634 & 1 & 60 & 53.9 & 165.8 & 151.5 & 151.8 & 362\\
1006 & 1 & 120 & 113.9 & 142.4 & 133.3 & 141.8 & 450\\
\textbf{1722} & \textbf{2} & \textbf{120} & \textbf{117.9} & \textbf{84.7} & \textbf{79.4} & \textbf{84.7} & \textbf{450}\\
2428 & 3 & 120 & 118.9 & 60.3 & 54.6 & 60.3 & 450\\
3130 & 4 & 120 & 119.3 & 46.8 & 42.6 & 47.0 & 450\\
\hline 
\end{tabular}
\tablefoot{$t$: time since the release of stars at the center of the host galaxy, in which the shell has reached its current radius calculated in the framework of the model of radial oscillations (Sect.~\ref{sub:rad_osc}); $n$: serial number of shell (Sect.~\ref{sub:TP}); $r_{\mathrm{s}}$: shell radius; $v_{\mathrm{s}}$: shell phase velocity according to the method described in Sect.~\ref{sub:edge}; $r_{\mathrm{TP}}$: galactocentric radius of current turning points of the stars at this time given by Eq.~(\ref{eq:Trn}); $v(r_{\mathrm{*,max}})$: radial velocity of stars at the shell edge; $v_{\mathrm{TP}}$: phase velocity of current turning point according Eq.~(\ref{eq:vTP}); $v_{\mathrm{c}}$: circular velocity at the shell edge radius. For parameters of the host galaxy, see Sect.~\ref{sec:param}. The shell that is used in Figs.~\ref{fig:zona}, \ref{fig:rmax}, \ref{fig:120app}, and \ref{fig:app-rez} is highlighted in bold.}
\end{table}

\begin{figure}
\centering{}
\resizebox{\hsize}{!}{\includegraphics[scale=0.3]{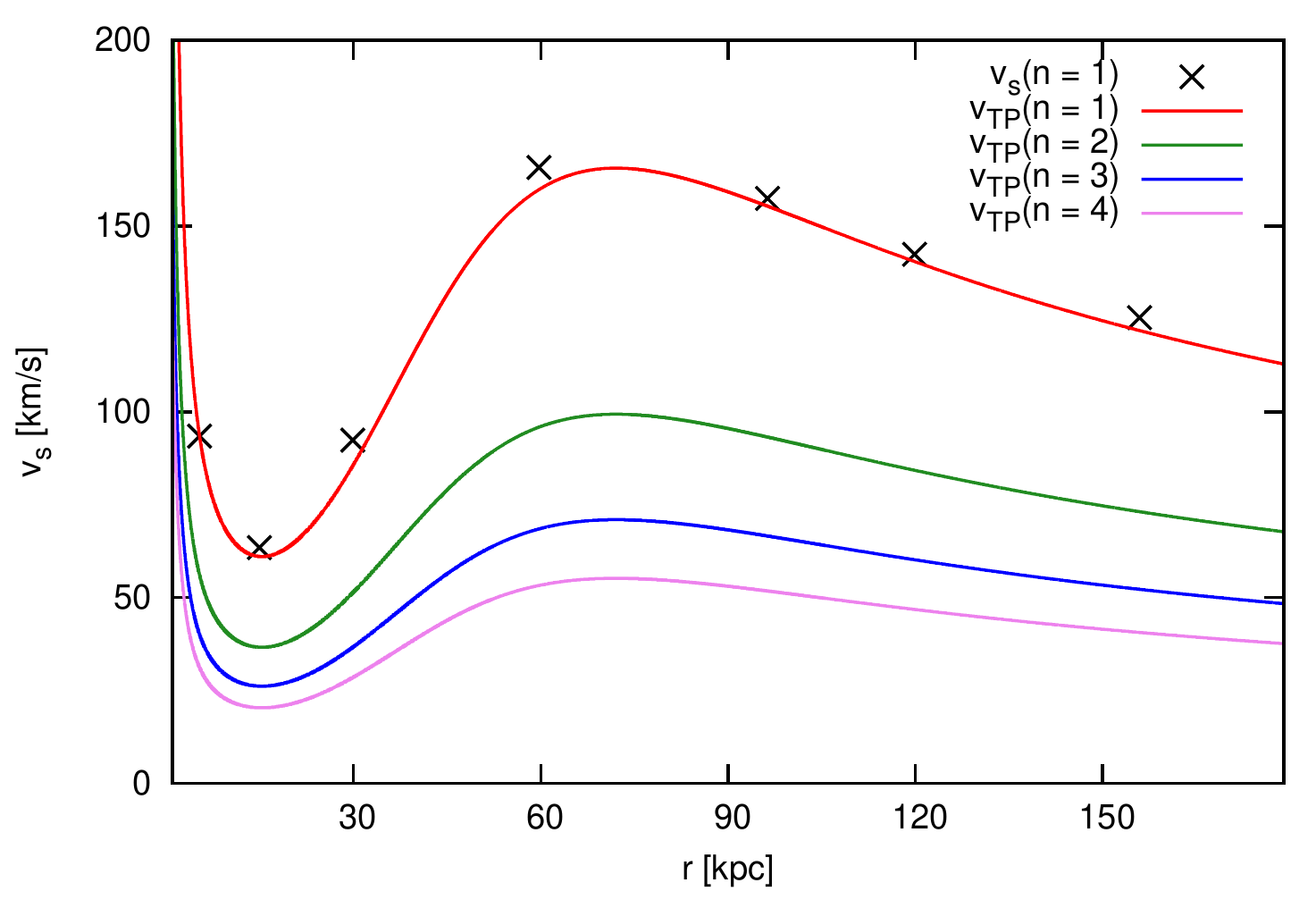}}
\caption{Dependence of the phase velocity of the turning points on the galactocentric radius for the first four shells according to Eq.~(\ref{eq:vTP}). For parameters of the host galaxy potential, see Sect.~\ref{sec:param}. Black crosses show the true velocity of the first shell calculated for several radii according to the method described in Sect.~\ref{sub:edge}. In fact, the turning point responsible for the current location of shell is not at the same radius as the shell edge at the same time, but the difference is small (Table~\ref{tab:param}). }
\label{fig:vs}
\end{figure}

The radial dependence of the phase velocity of the first four shells in the whole host galaxy is shown in Fig.~\ref{fig:vs}. Using Eq.~(\ref{eq:vTP}), we see that the velocity of each subsequent shell differs from the first one only by a factor of $3/(1+2n)$. The large interval of the galactocentric radii where the shell velocity increases is caused by the presence of the halo with a large scaling parameter. In fact, we do not show shell velocity, but the velocity of the turning points at the same radius. Nevertheless, these are sufficiently close. Black crosses show the true velocity of the first shell calculated for several radii according to the method described in Sect.~\ref{sub:edge}. For shells of higher $n$, these differences between the phase velocity of a shell and the corresponding turning point with consecutive serial numbers are even smaller.

\begin{figure}
\centering{}
\resizebox{\hsize}{!}{\includegraphics{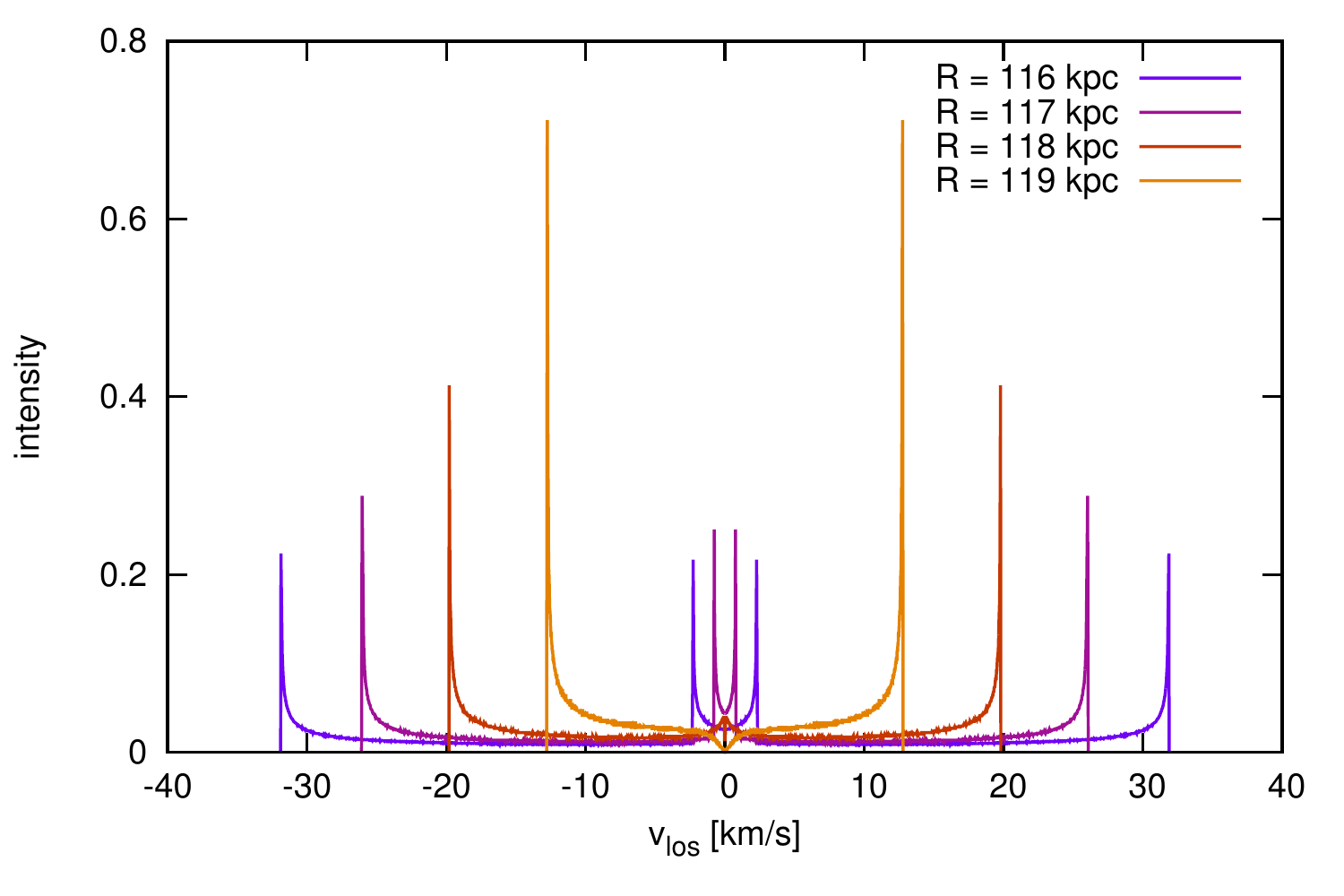}}
\caption{Evolution of the LOSVD near the shell edge for the second shell at $r_{\mathrm{s}}=120$\,kpc (parameters of the shell are highlighted in bold in Table~\ref{tab:param}) for the projected radius 116, 117, 118, and 119\,kpc in the framework of the model of radial oscillations (Sect.~\ref{sub:LOSVD-rad}). In this model, the current turning points of the shell particles are at $r_{\mathrm{TP}}=117.9$\,kpc. Beyond this radius, the inner maxima disappear. Profiles do not include stars of the host galaxy, which are not part of the shell system and are normalized so that the total flux equals one. For parameters of the host galaxy potential, see Sect.~\ref{sec:param}.}
\label{fig:zona}
\end{figure}

The radius of a stationary shell is the same as the radius of the apocenter of stars (as they all have the same energy), while the edge of a moving shell is at the radius, which is always slightly further from the center than the current turning points. This difference creates an intricate zone between the radius of the current turning points and the radius of the edge, where all the stars of a given shell move outwards. When the LOS radius from lower radii gets near to the turning points of the stars, the inner maxima of the LOSVD approach each other until they merge and finally disappear (Fig.~\ref{fig:zona}). We actually see a minimum in the middle of the LOSVD closer to the shell edge than the current turning points. The intricate zone is much larger for the first shell. For the shell radius of 120\,kpc in our host galaxy potential, it occupies 6\,kpc for the first shell, 2\,kpc for the second one, and less than one kpc for the fourth shell (Table~\ref{tab:param}). 

\section{Relating observables to circular and shell velocities} \label{sec:app}

The nonzero velocity of the shell complicates the kinematics of shells in two aspects mentioned above. Due to the energy difference between inward and outward particles at the same radius, the LOSVD peak is split into two and the shell edge is not at the radius of the current turning point, but slightly further from the center of the host galaxy. In this section, we describe the LOSVD of such a shell in the approximation of a locally constant galactic acceleration and shell velocity. In addition, we assume that the velocity of stars at the edge of the shell is equal to the phase velocity of the shell.


\subsection{Motion of a star in a shell system}\label{sub:star-app}

The galactocentric radius of the shell edge is a function of time, $r_{\mathrm{s}}(t)$, where $t=0$ is the moment of measurement and $r_{\mathrm{s}}(0)=r_{\mathrm{s0}}$ is the position of the shell edge at this time. We assume that the stars are on strictly radial orbits, that there is a locally constant value of the radial acceleration $a_{0}$ in the host galaxy potential and a locally constant velocity of the shell edge $v_{\mathrm{s}}$, and that the stars at the shell edge have the same velocity as the shell. The galactocentric radius of each star is at any time $r(t)$, while $t_{\mathrm{s}}$ is the time when the star could be found at the shell edge $r_{\mathrm{s}}(t_{\mathrm{s}})$. Then the equation of motion and the initial conditions for the star near a given shell radius are 
\begin{equation}
\frac{\mathrm{d}^{2}r(t)}{\mathrm{d}t^{2}} = a_{0},
\end{equation}
\begin{equation}
\left.\frac{\mathrm{d}r(t)}{\mathrm{d}t}\right|_{t=t_{\mathrm{s}}} = v_{\mathrm{s}},
\end{equation}
\begin{equation}
r(t_{\mathrm{s}}) = r_{\mathrm{s}}(t_{\mathrm{s}}) = v_{s}t_{\mathrm{s}}+r_{\mathrm{s0}}.
\end{equation}
The solution of these equations is
\begin{equation}
r(t)=a_{0}(t-t_{\mathrm{s}})^{2}/2+v_{\mathrm{s}}(t-t_{\mathrm{s}})+r_{\mathrm{s}}(t_{\mathrm{s}}),
\end{equation}
\begin{equation}
v(t)=v_{\mathrm{s}}+a_{0}(t-t_{\mathrm{s}}),
\end{equation}
and the actual position of the star $r(0)$ and its radial velocity $v(0)$ at time of measuring ($t=0$) are
\begin{equation}
r(0)=t_{\mathrm{s}}^{2}a_{0}/2+r_{\mathrm{s0}},
\end{equation}
\begin{equation}
v(0)=v_{\mathrm{s}}-a_{0}t_{\mathrm{s}}.
\end{equation}
Eliminating $t_{\mathrm{s}}$ from the two previous equations, we get
\begin{equation}
v(0)_{\pm}=v_{\mathrm{s}}\pm v_{\mathrm{c}}\sqrt{2\left(1-r(0)/r_{\mathrm{s0}}\right)},
\label{eq:v0}
\end{equation}
where $v_{\mathrm{c}}=\sqrt{-a_{0}r_{\mathrm{s0}}}$ is the circular velocity at the shell edge radius. 


\subsection{Approximative LOSVD}\label{sub:LOSVD-app}

The projection of the velocity given by Eq.~(\ref{eq:v0}) to the LOS at a projected radius $R$ will be 
\begin{equation}
\begin{array}{rcl}
v_{\mathrm{los}\pm} & = & \sqrt{1-R^{2}/\left(r\left(0\right)\right)^{2}}v(0)_{\pm} = \\
                    & = & \sqrt{1-R^{2}/\left(r\left(0\right)\right)^{2}}\left[v_{\mathrm{s}}\pm v_{\mathrm{c}}\sqrt{2\left(1-r(0)/r_{\mathrm{s0}}\right)}\right].
\end{array}
\label{eq:vlos}
\end{equation}
Using this expression, we can model the LOSVD at a given projected radius for a given shell. For the proper choice of a pair of values $v_{\mathrm{c}}$ and $v_{\mathrm{s}}$, we can find a match with observed and modeled peaks of the LOSVD.

\begin{figure}
\centering{}
\resizebox{\hsize}{!}{\includegraphics{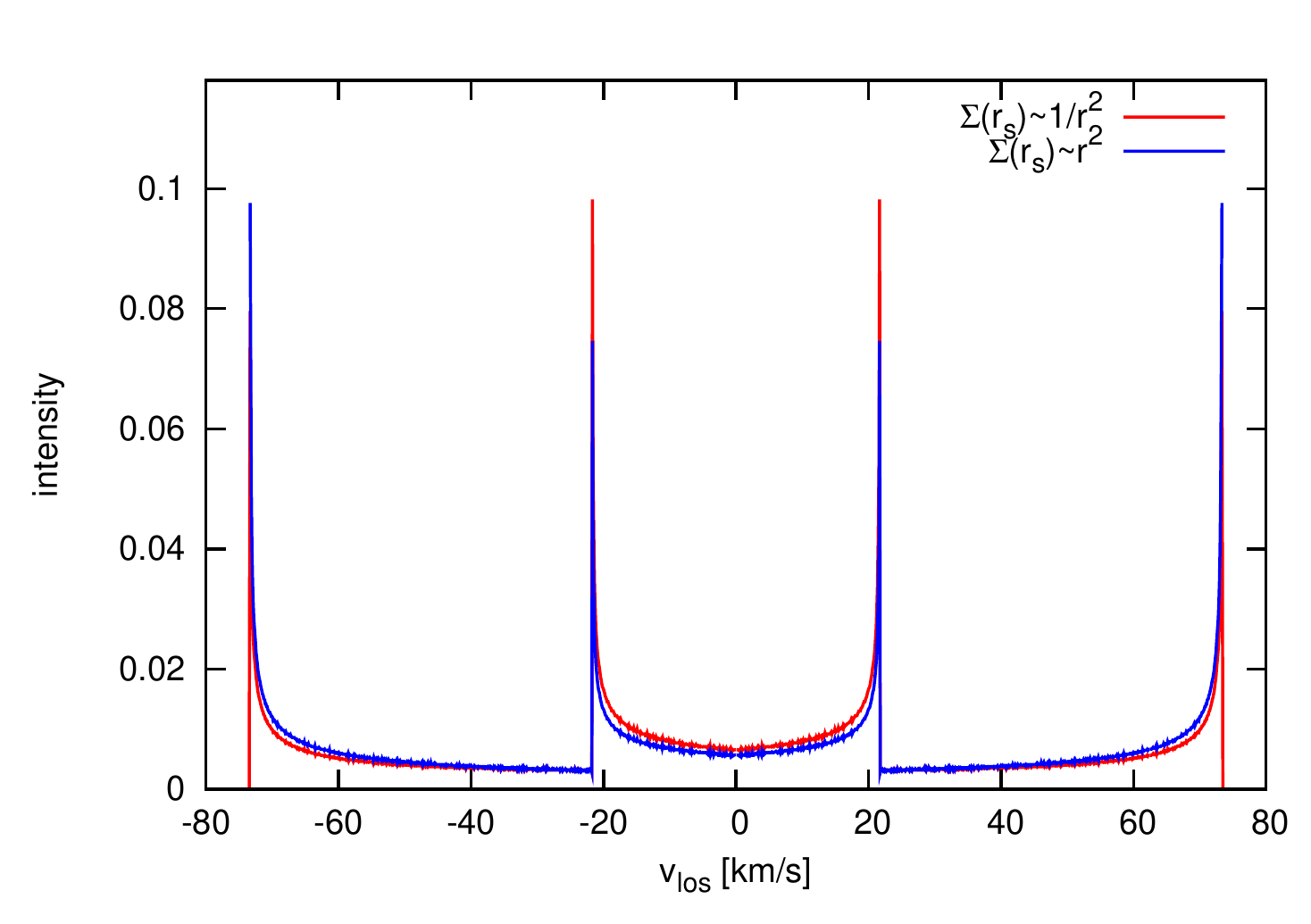}}
\caption{LOSVD of the second shell at $r_{\mathrm{s}}=120$\,kpc (parameters of the shell are highlighted in bold in Table~\ref{tab:param}) for the projected radius 108\,kpc in the framework of the model of radial oscillations (Sect.~\ref{sub:LOSVD-rad}), where the density at the surface of a sphere of shell edge radius $r_{\mathrm{s}}$ is $\Sigma_{\mathrm{sph}}(r_{\mathrm{s}}(t))\sim r_{\mathrm{s}}^{2}(t)$ for the blue curve and $\Sigma_{\mathrm{sph}}(r_{\mathrm{s}}(t))\sim1/r_{\mathrm{s}}^{2}(t)$ for the red one. The profile does not include stars of the host galaxy, which are not part of the shell system and are normalized, so that the total flux equals one.}
\label{fig:sigma}
\end{figure}

To model the LOSVD in both frameworks, the model of radial oscillations (Sect.~\ref{sub:LOSVD-rad}) and the approximative LOSVD by Eq.~(\ref{eq:vlos}), we have to add an assumption about the behavior of the shell brightness in time or in space (as the shell expands with time). This behavior depends on the parameters of the merger that has produced the shells. It is determined by the energy distribution of stars of the cannibalized galaxy in the instant of its decay in the center of the host galaxy. For simplicity, we choose the density at the surface of a sphere of shell edge radius $r_{\mathrm{s}}$ to be $\Sigma_{\mathrm{sph}}(r_{\mathrm{s}}(t))\sim1/r_{\mathrm{s}}^{2}(t)$, corresponding to a shell containing the same number of stars at each moment. The relation between $\Sigma_{\mathrm{sph}}(r_{\mathrm{s}}(t))$ and the projected surface density near the shell edge on the sky $\Sigma_{\mathrm{los}}(r_{\mathrm{s}}(t))$ is $\Sigma_{\mathrm{los}}(r_{\mathrm{s}}(t))\sim\sqrt{r_{\mathrm{s}}(t)}\Sigma_{\mathrm{sph}}(r_{\mathrm{s}}(t))$. It turns out that no reasonable choice of this function has an effect on the general characteristics of the LOSVD and the principles of formation that we describe in this paper. For illustration, we demonstrate the LOSVD of $\Sigma_{\mathrm{sph}}$ increasing as $r^2$ and $\Sigma_{\mathrm{sph}}$ decreasing as $1/r^2$ in Fig.~\ref{fig:sigma}. For the profiles shown, the ratio of the inner and outer peaks changes with the change of the $\Sigma_{\mathrm{sph}}$, but the peak positions are unaffected and the overall shape of the profile does not alter significantly. For shells that were created in a radial minor merger, we can expect a sharp rise in shell brightness near the center of the host galaxy, followed by an extensive area of its decrease. The fact that the main features of the LOSVD do not depend on the choice of $\Sigma_{\mathrm{sph}}$ means that our method of measuring the potential of shell galaxies is not sensitive to the details of the decay of the cannibalized galaxy.

\subsection{Radius of maximal LOS velocity}\label{sub:rvmax}

\citetalias{mk98} proved that near the edge of a stationary shell, $r_{s}$, the maximum intensity of the LOSVD is at the point where the maximal absolute value of the LOS velocity is. They also proved that the maximal absolute value of the LOS velocity $v_{\mathrm{los,max}}$ comes from stars at the galactocentric radius 
\begin{equation}
r_{v\mathrm{max}}=\frac{1}{2}(R+r_{\mathrm{s0}}),
\label{eq:1/2(R+rs)}
\end{equation}
at each projected radius $R$. 

\begin{figure}
\centering{}
\resizebox{\hsize}{!}{\includegraphics{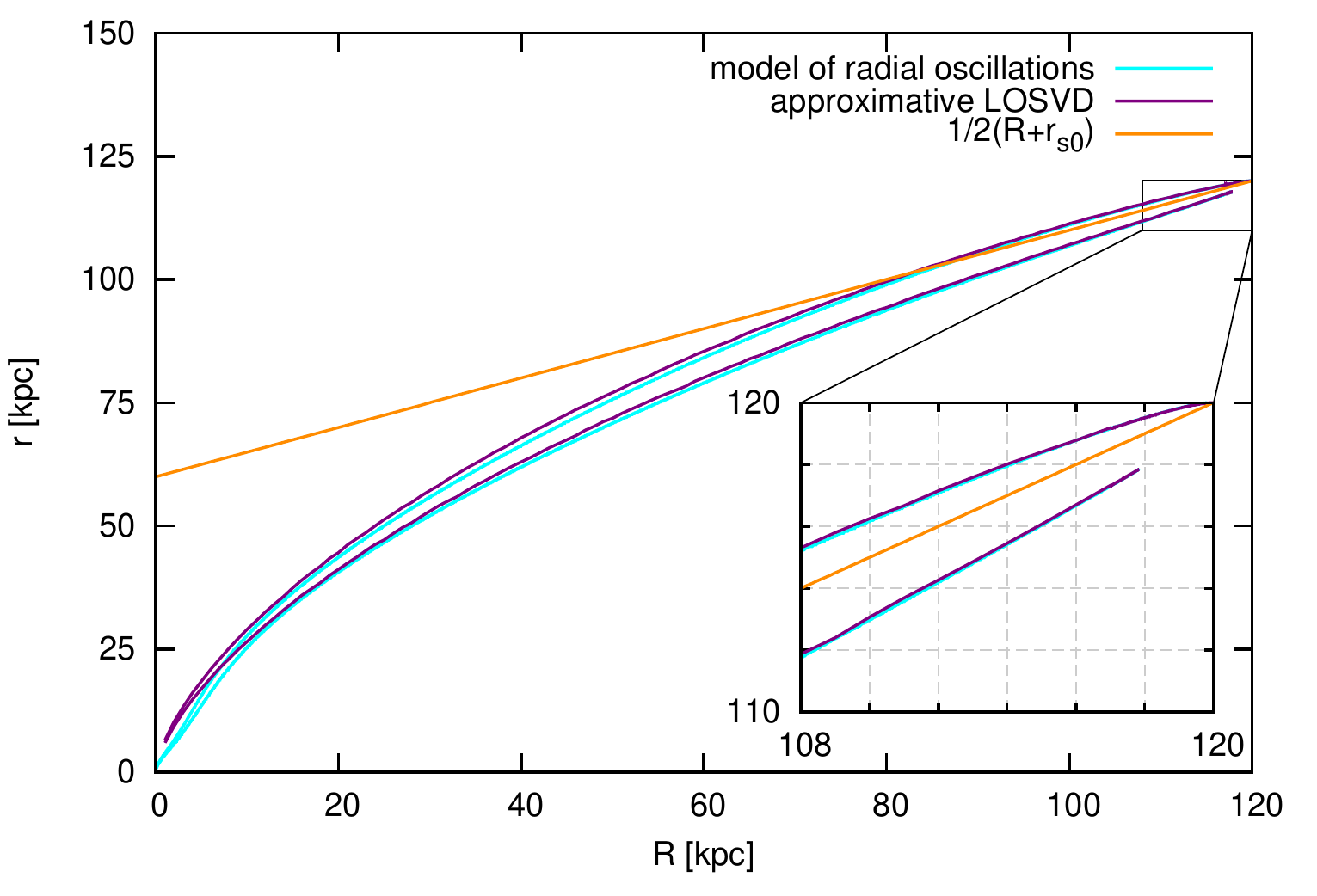}}
\caption{Galactocentric radii $r_{v\mathrm{max}}$ that contribute to the LOSVD the maximal velocities according to Eq.~(\ref{eq:1/2(R+rs)}), which was used in the derivation of the approximative maximal/minimal LOS velocities (Sect.~\ref{sec:Compars}, point~\ref{item:app-vmax})\,--\,orange curve, according to the approximative LOSVD (Sect.~\ref{sec:Compars}, point~\ref{item:app-LOSVD})\,--\,purple curves, and according to the model of radial oscillations (Sect.~\ref{sub:LOSVD-rad})\,--\,light blue curves for the second shell at\,120 kpc (parameters of the shell are highlighted in bold in Table~\ref{tab:param}). For parameters of the host galaxy potential, see Sect.~\ref{sec:param}.}
\label{fig:rmax}
\end{figure}

For a moving shell, analogous equations are significantly more complex and a similar relation cannot be easily proven. Nevertheless, when we apply both results of \citetalias{mk98} we can show in examples (Figs.~\ref{fig:120app}, \ref{fig:app-rez}, \ref{fig:vel.map}, and \ref{fig:rezy-sim}) that their use is valid, even for nonstationary shells. In the framework of the radial oscillations model (Sect.~\ref{sub:LOSVD-rad}), we have shown that the peaks of the LOSVD occur fairly close to the edges of distributions of inward and outward stars (Fig.~\ref{fig:anath}). The peaks are also near the edges of the LOSVD, if we divide the LOSVD into the contributions of the near and the far half of the galaxy as in Fig.~\ref{fig:anath} (b). The inner peak corresponds to inward-moving stars and the outer one to outward-moving ones. This approach is used in the equations in Sect.~\ref{sub:vlos,max}. The maximal LOS velocity corresponds to the outer peak and the minimal to the inner one. Reasons and justification for use of Eq.~(\ref{eq:1/2(R+rs)}) for $r_{v\mathrm{max}}$ are discussed in Sect.~\ref{sec:Compars}, point~\ref{item:app-vmax} (see also Fig.~\ref{fig:rmax}).

\subsection{Approximative maximal LOS velocity}\label{sub:vlos,max}

Using the results of \citetalias{mk98}, we derive an expression for the maxima/minima of the LOS velocity corresponding to locations of the LOSVD peaks in observable quantities (i.e., the maxima/minima of the LOS velocity, the projected radius, and the shell radius) 
by substituting $r_{v\mathrm{max}}$ given by Eq.~(\ref{eq:1/2(R+rs)}) for $r(0)$ in Eq.~(\ref{eq:vlos})
\begin{equation}
\begin{array}{rcl}
v_{\mathrm{los,max}\pm} \! &=& \! \left(v_{\mathrm{s}}\pm v_{\mathrm{c}}\sqrt{1-R/r_{\mathrm{s0}}}\right) \, \times \\
& & \times \sqrt{1-4\left(R/r_{\mathrm{s0}}\right)^{2} \left(1+R/r_{\mathrm{s0}}\right)^{-2}}.
\end{array}
\label{eq:vlos,max}
\end{equation}
For the measured locations of the LOSVD peaks $v_{\mathrm{los,max}+}$, $v_{\mathrm{los,max}-}$, projected radius $R$, and shell edge radius $r_{\mathrm{s0}}$, we can express the circular velocity $v_{\mathrm{c}}$ at the shell edge radius and the current shell velocity $v_{\mathrm{s}}$ by using inverse equations:
\begin{equation}
v_{\mathrm{c}}=\frac{\left|v_{\mathrm{los,max}+}-v_{\mathrm{los,max}-}\right|}{2\sqrt{\left(1-R/r_{\mathrm{s0}}\right)\left[1-4\left(R/r_{\mathrm{s0}}\right)^{2}\left(1+R/r_{\mathrm{s0}}\right)^{-2}\right]}}, 
\label{eq:vc,obs}
\end{equation}
\begin{equation}
v_{\mathrm{s}}=\frac{v_{\mathrm{los,max}+}+v_{\mathrm{los,max}-}}{2\sqrt{1-4\left(R/r_{\mathrm{s0}}\right)^{2}\left(1+R/r_{\mathrm{s0}}\right)^{-2}}}. 
\label{eq:vs,obs}
\end{equation}

Alternatively, the value of the circular velocity $v_{\mathrm{c}}$ at the shell edge radius could be inferred from measurements of positions of peaks at two or more different projected radii for the same shell: let $\bigtriangleup v_{\mathrm{los}}=v_{\mathrm{los,max}+}-v_{\mathrm{los,max}-}$, where $v_{\mathrm{los,max}\pm}$ satisfy Eq.~(\ref{eq:vlos,max}). Then, in the vicinity of the shell edge, 
\begin{equation}
\begin{array}{rcl}
\bigtriangleup v_{\mathrm{los}} &=& 2v_{\mathrm{c}}\sqrt{\left(R/r_{\mathrm{s0}}-1\right)\left[1-4\left(R/r_{\mathrm{s0}}\right)^{2}\left(1+R/r_{\mathrm{s0}}\right)^{-2}\right]} \approx \\ 
&\approx& 2(1-R/r_{\mathrm{s0}})v_{\mathrm{c}},
\end{array}
\end{equation}
and taking the derivative with respect to the projected radius
\begin{equation}
\frac{\mathrm{d}\!\bigtriangleup\! v_{\mathrm{los}}}{\mathrm{d}R}=-2\frac{v_{\mathrm{c}}}{r_{\mathrm{s0}}},\label{eq:sklon}
\end{equation}
which happens to be the same expression as equation (7) in \citetalias{mk98}. Nevertheless, in \citetalias{mk98}, $\bigtriangleup v_{\mathrm{los}}$ is the distance between the two LOSVD intensity maxima of a stationary shell, whereas in our framework, it is the distance between the outer peak for positive velocities and the inner peak for negative velocities or vice versa. This equation allows us to measure the circular velocity in shell galaxies using the slope of the LOSVD intensity maxima in the $R \times v_{\mathrm{los}}$ diagram.


\section{Comparison of models} \label{sec:Compars}

In this section, we compare three different approaches to the theoretical calculation of the maximal/minimal LOS velocities, which are equivalent to the positions of LOSVD peaks:

\begin{enumerate}

\item Using the model of radial oscillations as described in Sect.~\ref{sub:LOSVD-rad} (these results are plotted with light blue curves in relevant figures). This model requires thorough knowledge of the potential of the host galaxy, obviously unavailable for real galaxies.

\item Using the approximative LOSVD (purple curves). For the given shell at the chosen projected radius, Eq.~(\ref{eq:vlos}) is a function of only two parameters, the circular velocity $v_{\mathrm{c}}$ at the shell edge radius and the current shell velocity $v_{\mathrm{s}}$. Assuming a behavior of shell brightness as a function of the shell radius, Eq.~(\ref{eq:vlos}) allows us to plot the whole LOSVD (Sect.~\ref{sub:LOSVD-app}). However, computing the LOSVD and the peaks' positions requires a numerical approach in this framework.
\label{item:app-LOSVD}

\item Using the approximative maximal LOS velocities (orange curves). Eq.~(\ref{eq:vlos,max}) supplies the positions of the peaks directly. It differs from the previous approximation in the assumption about the galactocentric radius $r_{v\mathrm{max}}$, from which comes the contribution to the LOSVD at the maximal speed. The assumption is that $r_{v\mathrm{max}}$  is given by Eq.~(\ref{eq:1/2(R+rs)}), which was derived by \citetalias{mk98} for a stationary shell. This equation is actually only very approximate, but allows us to analytically invert Eq.~(\ref{eq:vlos,max}) to obtain formulae for the calculation of $v_{\mathrm{c}}$ and $v_{\mathrm{s}}$ from the measured peak positions in the spectrum of the shell galaxy near the shell edge (Eqs.~(\ref{eq:vc,obs}) and (\ref{eq:vs,obs})). For a moving shell, we could not derive a more accurate formula for $r_{v\mathrm{max}}$ that would be simple enough to make the calculation of $v_{\mathrm{c}}$ and $v_{\mathrm{s}}$ feasible.
\label{item:app-vmax}
\end{enumerate}

Fig.~\ref{fig:rmax} shows a comparison of the radii that contribute to the LOSVD at the maximal velocities according to all three approaches. For the first two methods, the radius corresponding to the inner maxima of the LOSVD (which are the maxima created by the inward stars) is lower than that for the outer maxima, whereas Eq.~(\ref{eq:1/2(R+rs)}) assumes the same $r_{v\mathrm{max}}$ for both inward and outward stars.

\begin{figure}
\centering{}
\resizebox{\hsize}{!}{\includegraphics{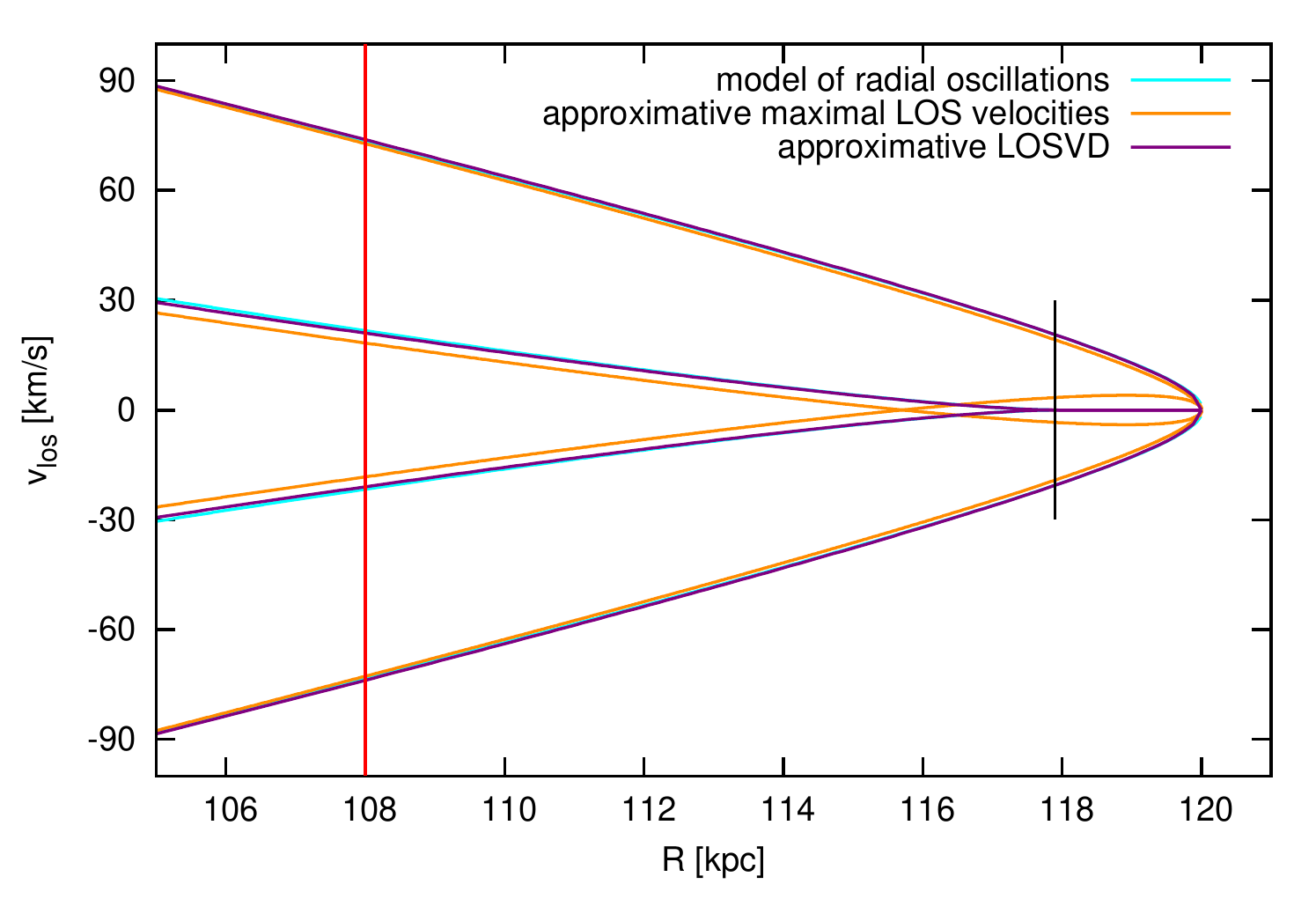}}
\caption{LOSVD peak locations for the second shell at the radius of 120\,kpc (parameters of the shell are highlighted in bold in Table~\ref{tab:param}) according to the approximative maximal LOS velocities (Sect.~\ref{sec:Compars}, point~\ref{item:app-vmax}) given by Eq.~(\ref{eq:vlos,max}) (orange curves); the approximative LOSVD (Sect.~\ref{sec:Compars}, point~\ref{item:app-LOSVD}) given by Eq.~(\ref{eq:vlos}) (purple curves); and the model of radial oscillations (Sect.~\ref{sub:LOSVD-rad}) (light blue curves almost merged with the purple ones). The red line shows the position of the LOSVD from Fig.~\ref{fig:app-rez}, the black one shows the position of the current turning points. For parameters of the host galaxy potential, see Sect.~\ref{sec:param}.}
\label{fig:120app}
\end{figure}

Fig.~\ref{fig:120app} shows locations of the LOSVD peaks for the second shell at the radius of 120\,kpc near the shell edge radius. The purple curve is calculated using the approximative LOSVD (Sect.~\ref{sec:Compars}, point~\ref{item:app-LOSVD}) given by Eq.~(\ref{eq:vlos}), into which we inserted the velocity of the second shell according to the model of radial oscillations and the circular velocity in the potential of the host galaxy (see Sect.~\ref{sec:param} for parameters of the potential). The purple curve does not differ significantly from the light blue curve calculated in the model of radial oscillations (Sect.~\ref{sub:LOSVD-rad}). The more important deviations in the orange curve of the approximative maximal LOS velocities (Sect.~\ref{sec:Compars}, point~\ref{item:app-vmax}) given by Eq.~(\ref{eq:vlos,max}), are caused by using Eq.~(\ref{eq:1/2(R+rs)}) for $r_{v\mathrm{max}}$. With this assumption, approximative maximal LOS velocities (the orange curve) predict that around the zone between the current turning point and the shell edge,  the inner peaks change signs. This means that for the part of the galaxy closer to the observer, both inner and outer peaks will fall into negative values of the LOS velocity and vice versa. However, from the model of the radial oscillations we know that the signal from the inner peak in a given (near or far) part of the galaxy is always zero or has the opposite sign to that of the outer peak.

\begin{figure}
\centering{}
\resizebox{\hsize}{!}{\includegraphics{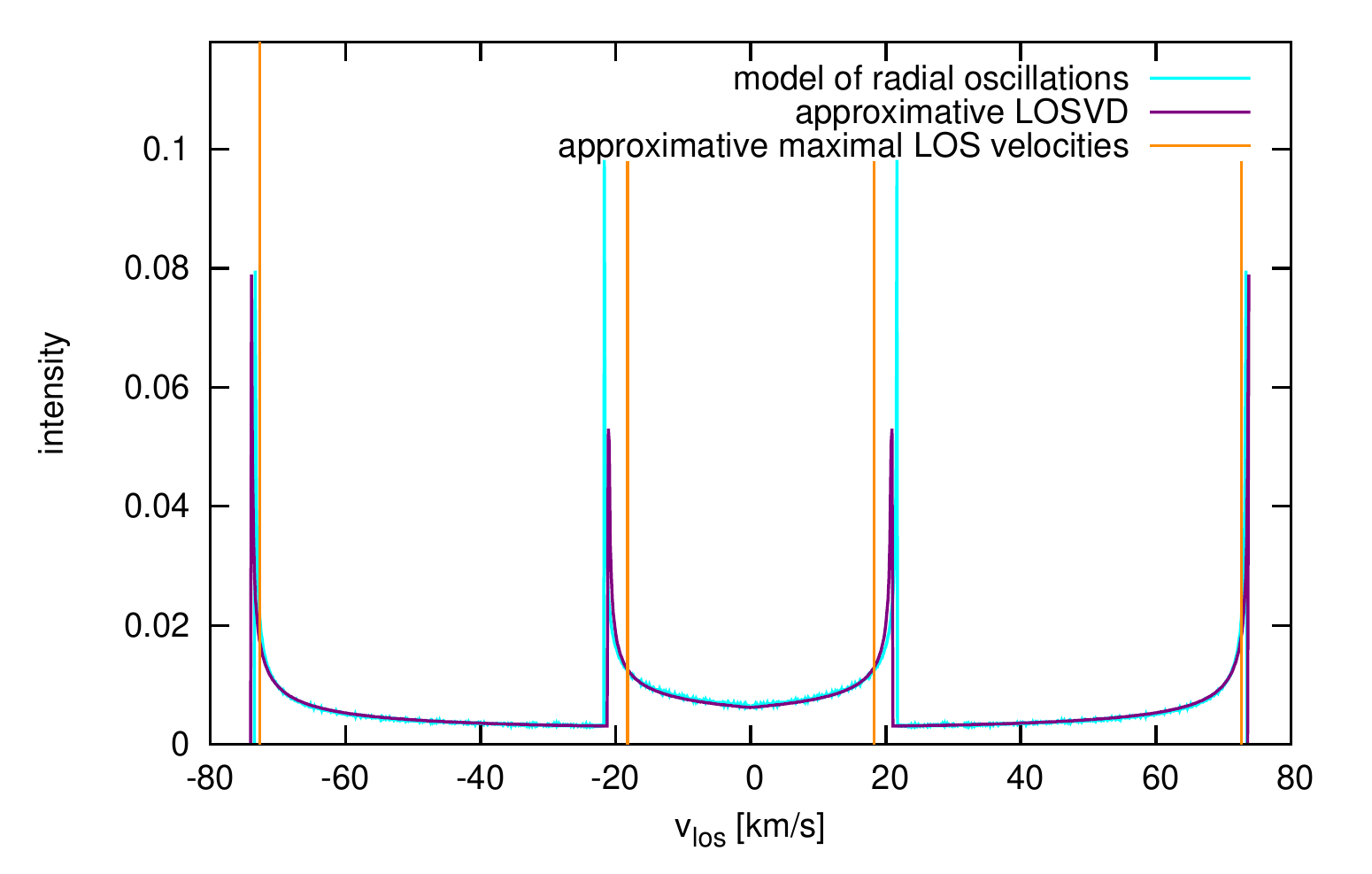}}
\caption{LOSVD of the second shell at $r_{\mathrm{s}}=120$\,kpc (parameters of the shell are highlighted in bold in Table~\ref{tab:param}) for the projected radius $R=0.9r_{\mathrm{s}}=108$\,kpc according to the approximative LOSVD (Sect.~\ref{sec:Compars}, point~\ref{item:app-LOSVD}) given by Eq.~(\ref{eq:vlos}) (purple curve) and the model of radial oscillations (Sect.~\ref{sub:LOSVD-rad}) (light blue curve almost merged with the purple one). Locations of peaks as given by the approximative maximal LOS velocities (Sect.~\ref{sec:Compars}, point~\ref{item:app-vmax}) given by Eq.~(\ref{eq:vlos,max}) are plotted with orange lines. Profiles do not include stars of the host galaxy that are not part of the shell system and are normalized, so that the total flux equals to one. For parameters of the host galaxy potential see Sect.~\ref{sec:param}.}
\label{fig:app-rez}
\end{figure}

The model of the radial oscillations and the approximative LOSVD given by Eq.~(\ref{eq:vlos}) were also used to construct the LOSVD for the second shell located at 120\,kpc, at the projected radius of 108\,kpc in Fig.~\ref{fig:app-rez}. The graph also shows the locations of the peaks using the approximative maximal LOS velocities given by Eq.~(\ref{eq:vlos,max}).

\section{Test-particle simulation of the merger}\label{sec:N-Simulations}

We performed a simplified simulation of formation of shells in a radial galactic minor merge. Both merging galaxies are represented by smooth potential. Millions of test particles were generated so that they follow the distribution function of the cannibalized galaxy at the beginning of the simulation. The particles then move according to the sum of the gravitational potentials of both galaxies. When the centers of the galaxies pass through each other, the potential of the cannibalized galaxy is suddenly switched off and the particles continue to move only in the fixed potential of the host galaxy. We use the simulation to demonstrate the validity of our methods of recovering the parameters of the host galaxy potential by measuring\,\footnote{Here and in the rest of this section, the data measured are the output of our simulation.} the positions of the peaks in the spectral lines. In all cases, we look at the galaxy from the view perpendicular to the axis of collision, so that the cannibalized galaxy originally flew in from the right.


\subsection{Parameters of the simulation}\label{sec:param}

The potential of the host galaxy is modeled as a double Plummer sphere with respective masses $M_{*}=2\times10^{11}$\,M$_{\sun}$ and $M_{\mathrm{DM}}=1.2\times10^{13}$\,M$_{\sun}$, and Plummer radii $b_{*}=5$\,kpc and $b_{\mathrm{DM}}=100$\,kpc for the luminous component and the dark halo, respectively. This model has properties consistent with observed massive early-type (and even shell) galaxies \citep{auger10,nagino09,fukazawa06}. The potential of the cannibalized galaxy is chosen to be a single Plummer sphere with the total mass $M=2\times10^{10}$\,M$_{\sun}$ and Plummer radius $b_{*}=2$\,kpc. 

\begin{figure}
\centering{}
\resizebox{\hsize}{!}{\includegraphics{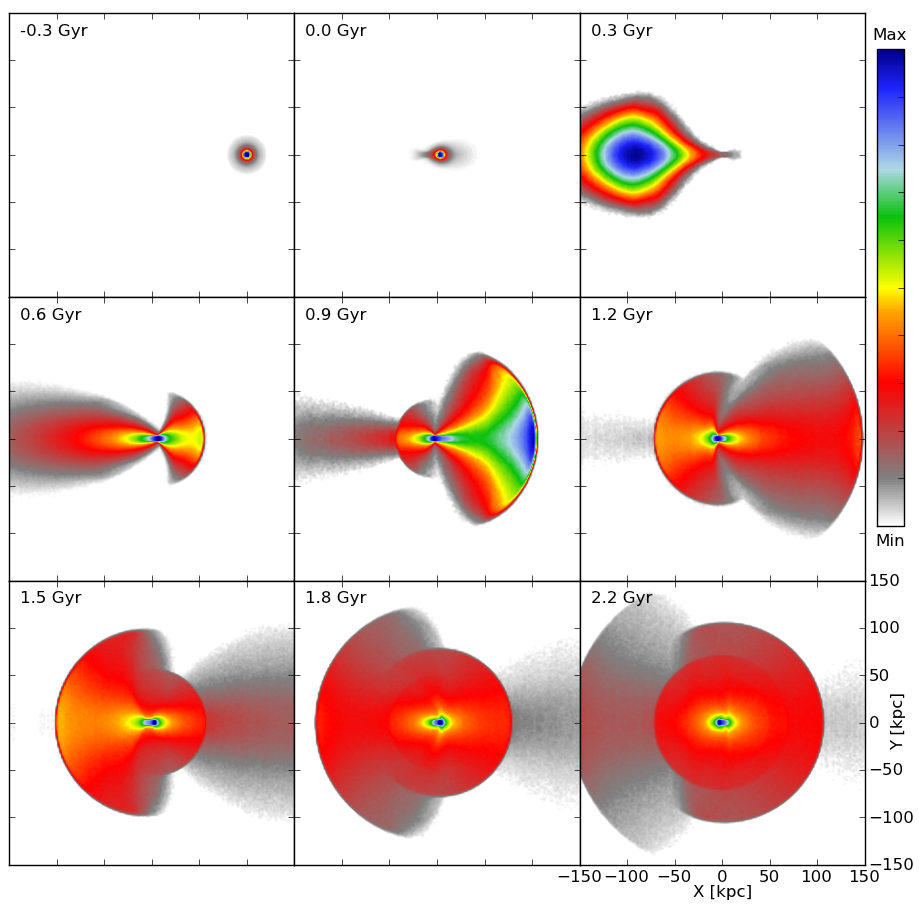}}
\caption{Snapshots from our test-particle simulation of the radial minor merger, leading to the formation of shells. Each panel covers 300$\times$300\,kpc and is centered on the host galaxy. Only the surface density of particles originally belonging to the satellite galaxy is displayed. The density scale varies between frames, so that the respective range of densities is optimally covered.}
\label{fig:movie}
\end{figure}

The cannibalized galaxy is released from rest at a distance of 100\,kpc from the center of the host galaxy. When it reaches the center of the host galaxy in 306.4\,Myr, its potential is switched off and its particles begin to oscillate freely in the host galaxy. The shells start appearing visibly from about 50\,kpc of galactocentric distance and disappear at around 200\,kpc, as there are very few particles with apocenters outside these radii (Fig.~\ref{fig:movie}).

\subsection{Comparison of the simulation with models}\label{sec:sim-mod}

In the simulations, some of the assumptions that we used earlier (Sect.~\ref{sub:rad_osc}) are not fulfilled. First, the particles do not move radially, but on more general trajectories, which are, even in the case of a radial merger, nevertheless very eccentric. Second, not all the particles are released from the cannibalized galaxy right in the center of the host galaxy; when the potential is switched off, the particles are located in the broad surroundings of the center and some are even released before the decay of the galaxy. These effects cause a smearing of the kinematical imprint of shells, as the turning points are not at a sharply defined radius, but rather in some interval of radii for a given time. 

\begin{figure}
\centering{}
\resizebox{\hsize}{!}{\includegraphics{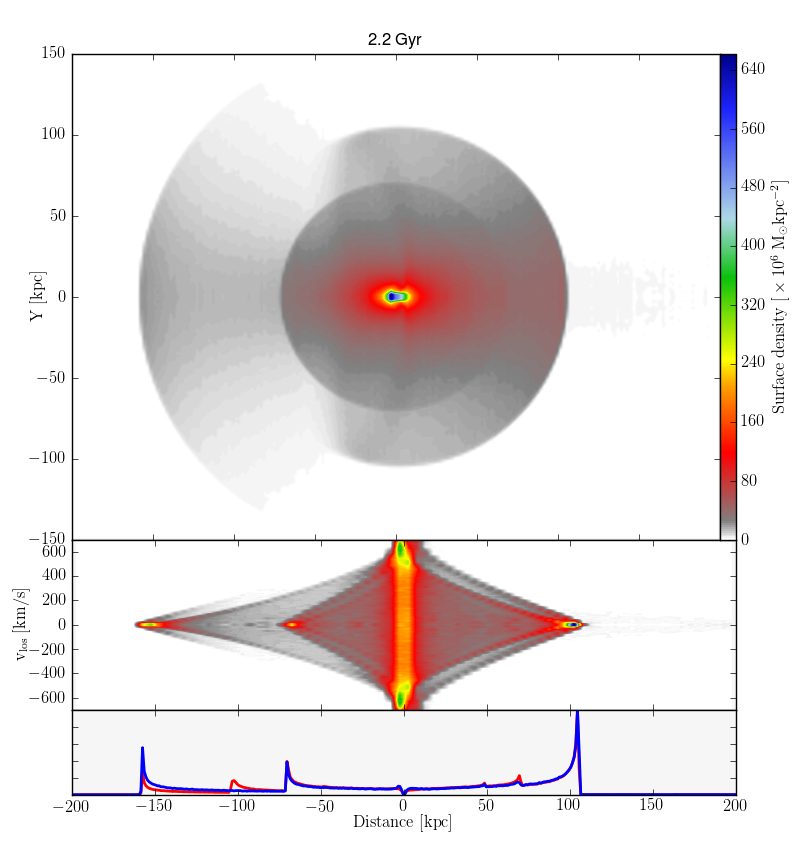}}
\caption{Simulated shell structure 2.2\,Gyr after the decay of the cannibalized galaxy. Only the particles originally belonging to the cannibalized galaxy are taken into account. Top: surface density map; middle: the LOSVD density map of particles in the $\pm1$\,kpc band around the collision axis; bottom: histogram of galactocentric distances of particles. The angle between the radial position vector of the particle and the $x$-axis (the collision axis) is less than 90$\degr$ for the blue curve and less than 45$\degr$ for the red curve. The horizontal axis corresponds to the projected distance $X$ in the upper panel, to the projected radius $R$ in the middle panel, and to the galactocentric distance $r$ in the lower panel.}
\label{fig:2200Myr}
\end{figure}

The model of radial oscillations presented in Sect.~\ref{sub:rad_osc} predicts that 2.2\,Gyr after the decay of the cannibalized galaxy (Fig.~\ref{fig:2200Myr}), five outermost shells should lie at the radii of 257.3, $-$157.8, 105.1, $-$70.5, and 48.8\,kpc. The negative radii refer to the shell being on the opposite side of the host galaxy with respect to the direction from which the cannibalized galaxy flew in. These radii agree well with the radii of the shells measured in the simulation 2.2\,Gyr after the decay of the cannibalized galaxy. In the simulation, the first shell at 257.4\,kpc is composed of only a few particles, and therefore we will not consider it (its parameters are listed in Table~\ref{tab:param-sim} for completeness). Thus, the outermost relevant shell in the system lies at $-$157.8\,kpc and has a serial number $n=2$. Also, the shell at 48.8\,kpc suffers from lack of particles, but we will include it nevertheless.

\begin{table}
\centering
\caption{Parameters of the shells in a simulation 2.2\,Gyr after the decay of the cannibalized galaxy.}
\begin{tabular}{cccccc}
\hline \hline 
$r_{\mathrm{s}}$ & $n$ & $r_{\mathrm{TP,model}}$ & $v_{\mathrm{s,sim}}$ & $v_{\mathrm{s,model}}$ & $v\mathrm{_{c,model}}$\\
kpc &   & kpc & km$/$s & km$/$s & km$/$s\\
\hline 
48.8 & 5 & 48.5 & 38.7$\pm$2.1 & 38.7 & 326\\
$-$70.6 & 4 & $-$69.9 & 59.8$\pm$1.6 & 54.3 & 390\\
105.0 & 3 & 103.9 & 68.1$\pm$1.9 & 63.5 & 441\\
$-$157.8 & 2 & $-$155.7 & 74.3$\pm$1.2 & 72.4 & 450\\
257.4 & 1 & 251.0 & 97.5$\pm$1.4 & 95.7 & 406\\
\hline 
\end{tabular}
\tablefoot{The values of $r_{\mathrm{TP,model}}$ and $v_{\mathrm{s,model}}$ are calculated for the shell position $r_{\mathrm{s}}$ and its corresponding serial number $n$ according to the model of radial oscillations (Sect.~\ref{sub:rad_osc}). The shell velocity $v_{\mathrm{s,sim}}$ is derived from 20 positions between the times 2.49--2.51\,Gyr for each shell. The value $v\mathrm{_{c,model}}$ corresponds to the circular velocity at the shell edge radius $r_{\mathrm{s}}$ for the chosen potential of the host galaxy (Sect.~\ref{sec:param}).}
\label{tab:param-sim}
\end{table}

\begin{figure}
\centering{}
\resizebox{\hsize}{!}{\includegraphics{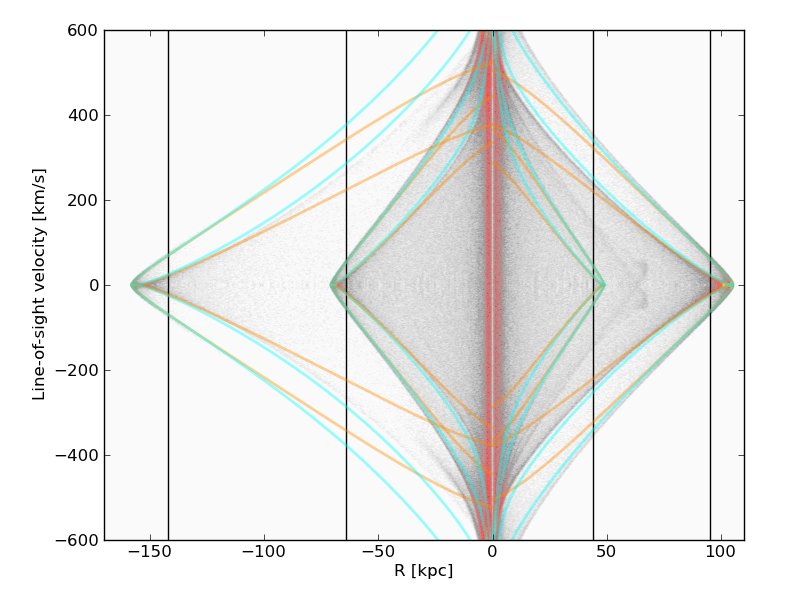}}
\caption{LOSVD map of the simulated shell structure 2.2\,Gyr after the decay of the cannibalized galaxy (middle panel in Fig.~\ref{fig:2200Myr}). Light blue curves show locations of the maxima according to the model of radial oscillations (Sect.~\ref{sub:LOSVD-rad}) for shell radius $r_{\mathrm{s}}$, corresponding serial number $n$, and the known potential of the host galaxy (Sect.~\ref{sec:param}). Orange curves are derived from the approximative maximal LOS velocities (Sect.~\ref{sec:Compars}, point~\ref{item:app-vmax}) given by Eq.~(\ref{eq:vlos,max}) for $r_{\mathrm{s}}$, $v_{\mathrm{s,model}}$, and $v\mathrm{_{c,model}}$. Parameters of the shells are shown in Table~\ref{tab:param-sim}. Black lines mark the location at $0.9r_{\mathrm{s}}$ for each shell. The LOSVD for these locations are shown in Fig.~\ref{fig:rezy-sim}. The map includes only stars originally belonging to the cannibalized galaxy.}
\label{fig:vel.map}
\end{figure}

Fig.~\ref{fig:vel.map} shows the comparison between the LOSVD in the simulation, the peaks of the LOSVD computed in the model of radial oscillations (light blue curves), and the approximative maximal LOS velocities\,--\,Eq.~(\ref{eq:vlos,max}) (orange curves). To evaluate the approximative maximal LOS velocities, we obtained the shell velocity $v_{\mathrm{\mathrm{s,model}}}$ from the model of radial oscillations (Sect.~\ref{sub:rad_osc}) for the respective serial number $n$ of the shell and circular velocity $v\mathrm{_{c,model}}$ at the shell edge radius, using our knowledge of the potential of the host galaxy (see Sect.~\ref{sec:param} for parameters of the potential). The values of all the respective shell quantities are listed in Table~\ref{tab:param-sim}. 

\begin{figure*}
\centering{}
\resizebox{\hsize}{!}{\includegraphics{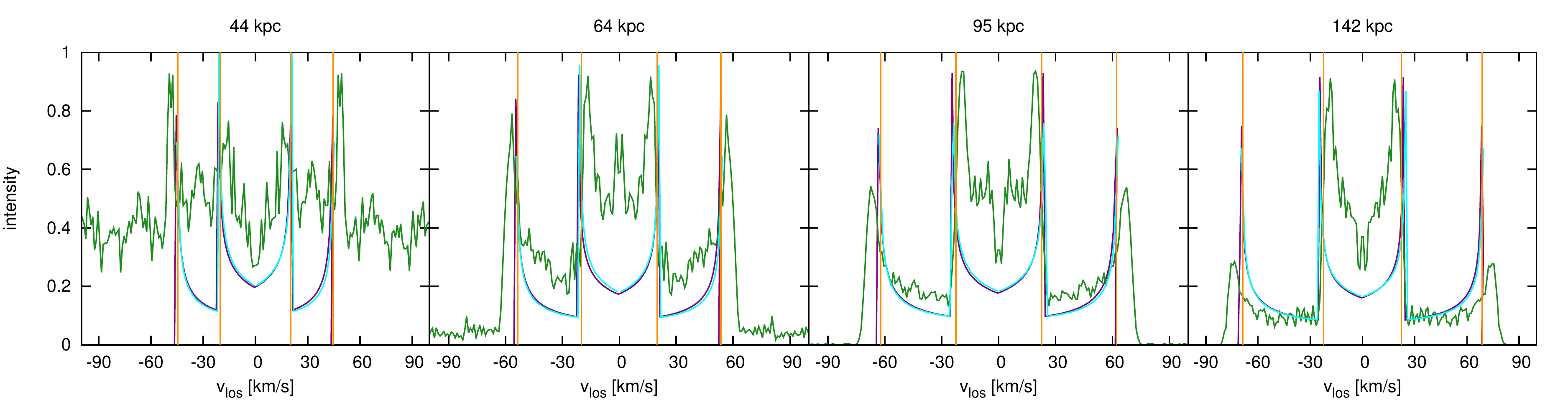}}
\caption{LOSVDs of four shells at projected radii $0.9r_{\mathrm{s}}$ (indicated as the title of each plot) 2.2\,Gyr after the decay of the cannibalized galaxy (parameters of the shells are shown in Table~\ref{tab:param-sim}). The simulated data are shown in green, the LOSVDs according to the approximative LOSVD (Sect.~\ref{sec:Compars}, point~\ref{item:app-LOSVD}) given by Eq.~(\ref{eq:vlos}) in purple, and LOSVDs according to the model of radial oscillations (Sect.~\ref{sub:LOSVD-rad}) in light blue. The graph also shows the locations of the peaks using the approximative maximal LOS velocities (Sect.~\ref{sec:Compars}, point~\ref{item:app-vmax}) given by Eq.~(\ref{eq:vlos,max}) by orange lines. Profiles do not include stars of the host galaxy, which are not part of the shell system. The theoretical profiles are scaled so that the intensity of their highest peak approximately agrees with the highest peak of the simulated data. Intensity is given in relative units, so maxima of the profiles have values of about 0.9.}
\label{fig:rezy-sim}
\end{figure*}

Fig.~\ref{fig:vel.map} also shows the locations that correspond to the radii of $0.9r_{\mathrm{s}}$ for each individual shell (black lines). 
The LOSVD for these locations is shown in Fig.~\ref{fig:rezy-sim}. The positions of simulated LOSVD peaks largely agree with three theoretical approaches described in Sect.~\ref{sec:Compars}. 

\begin{table*}
\centering
\caption{Circular velocity at the shell edge radius $r_{\mathrm{s}}$ derived from the measurement of the simulated data 2.2\,Gyr after the decay of the cannibalized galaxy.}
\begin{tabular}{cccccccccc}
\hline \hline
 \\[-0.3cm]
$r_{\mathrm{s}}$ & $v\mathrm{_{c,model}}$ & $N_{\mathrm{data}}$ & $N_{\mathrm{data}}^{\mathrm{SS}}$ & $v_{\mathrm{c,eq(\ref{eq:vc,obs})}}$ & $v_{\mathrm{c,eq(\ref{eq:vc,obs})}}^{\mathrm{SS}}$ & $v_{\mathrm{c,slope}}$ & $v_{\mathrm{c,slope}}^{\mathrm{SS}}$ & $v_{\mathrm{c,fit}}$ & $v_{\mathrm{c,slope(MK98)}}$\\
kpc & km$/$s &  &  & km$/$s & km$/$s & km$/$s & km$/$s & km$/$s & km$/$s\\
\hline 
48.8     & 326 & 5  & 4  & 346$\pm$130 & 340$\pm$94 & 322$\pm$19 & 314$\pm$32 & 318$\pm$51 & 449$\pm$26\\
$-$70.6  & 390 & 7  & 5  & 394$\pm$85  & 390$\pm$53 & 391$\pm$5  & 392$\pm$11 & 368$\pm$60 & 570$\pm$23\\
105.0    & 441 & 11 & 8  & 478$\pm$144 & 452$\pm$64 & 440$\pm$5  & 447$\pm$7  & 427$\pm$28 & 632$\pm$9\\
$-$157.8 & 450 & 15 & 10 & 497$\pm$236 & 472$\pm$79 & 462$\pm$8  & 484$\pm$14 & 460$\pm$32 & 671$\pm$11\\
\hline 
\end{tabular}
\tablefoot{$r_{\mathrm{s}}$ and $v\mathrm{_{c,model}}$ have the same meaning as in Table~\ref{tab:param-sim}. $N_{\mathrm{data}}$: number of measurements for each shell; $v_{\mathrm{c,eq(\ref{eq:vc,obs})}}$: the mean of values derived from the approximative maximal LOS velocities given by Eq.~(\ref{eq:vc,obs}) with its mean square deviation; $v_{\mathrm{c,slope}}$: a value derived from linear regression using the slope of the LOSVD intensity maxima given by Eq.~(\ref{eq:sklon}) and its standard error (see also Fig.~\ref{fig:vccomp}); $v_{\mathrm{c,fit}}$: a value derived by fitting a pair of $v_{\mathrm{c}}$ and $v_{\mathrm{s}}$ in the approximative LOSVD given by Eq.~(\ref{eq:vlos}) (Sect.~\ref{sec:Compars}, point~\ref{item:app-LOSVD} and Fig.~\ref{fig:minfit}); $v_{\mathrm{c,slope(MK98)}}$: the mean of values derived from the slope of the LOSVD intensity maxima given by Eq.~(\ref{eq:sklon}) with its standard error (see also Fig.~\ref{fig:vccomp}). In the equation, however, $\bigtriangleup v_{\mathrm{los}}$ is substituted with the distance between the two outer peaks of the LOSVD intensity maxima in order to mimic the measurement as originally proposed by \citetalias{mk98} for double-peaked profile. The quantities with the superscript SS correspond to the subsample, where only measurements with two discernible inner peaks in the LOSVD are used.}
\label{tab:param-vc}
\end{table*}

\begin{table*}
\centering
\caption{Velocity of the shell at the radius $r_{\mathrm{s}}$ derived from the measurement of the simulated data 2.2\,Gyr after the decay of the cannibalized galaxy.}
\begin{tabular}{cccccccccc}
\hline \hline
 \\[-0.3cm]
$r_{\mathrm{s}}$ & $v_{\mathrm{s,model}}$ & $N_{\mathrm{data}}$ & $N_{\mathrm{data}}^{\mathrm{SS}}$ & $v_{\mathrm{s,sim}}$ & $v_{\mathrm{s,eq(\ref{eq:vs,obs})}}$ & $v_{\mathrm{s,eq(\ref{eq:vs,obs})}}^{\mathrm{SS}}$ & $v_{\mathrm{s,eq(\ref{eq:vs-vc})-slope}}$ & $v_{\mathrm{s,eq(\ref{eq:vs-vc})-slope}}^{\mathrm{SS}}$ & $v_{\mathrm{s,fit}}$\\
kpc & km$/$s &  &  & km$/$s & km$/$s & km$/$s & km$/$s & km$/$s & km$/$s\\
\hline
48.8     & 38.7 & 5  & 4  & 38.7$\pm$2.1 & 50.7$\pm$2.3 & 51.7$\pm$1.1 & 44.2$\pm$6.5  & 44.9$\pm$6.3 & 53$\pm$16\\
$-$70.6  & 54.3 & 7  & 5  & 59.8$\pm$1.6 & 60.8$\pm$9.8 & 65.6$\pm$2.0 & 60.7$\pm$10.8 & 66.0$\pm$2.9 & 66$\pm$19\\
105.0    & 63.5 & 11 & 8  & 68.1$\pm$1.9 & 74.8$\pm$4.6 & 76.5$\pm$1.4 & 68.0$\pm$8.9  & 71.3$\pm$2.5 & 79$\pm$9\\
$-$157.8 & 72.4 & 15 & 10 & 74.3$\pm$1.2 & 84.4$\pm$5.4 & 86.7$\pm$2.0 & 78.7$\pm$10.5 & 82.$\pm$3.5  & 85$\pm$14\\

\hline
\end{tabular}
\tablefoot{$r_{\mathrm{s}}$, $v_{\mathrm{s,model}}$, and $v_{\mathrm{s,sim}}$ have the same meaning as in Table~\ref{tab:param-sim}. $N_{\mathrm{data}}$: number of measurements for each shell; $v_{\mathrm{s,eq(\ref{eq:vs,obs})}}$: the mean of values derived from the approximative maximal LOS velocities given by Eq.~(\ref{eq:vs,obs}) with its mean square deviation; $v_{\mathrm{s,eq(\ref{eq:vs-vc})-slope}}$: the mean of values derived from the hybrid relation given by Eq.~(\ref{eq:vs-vc}) with its mean square deviation (see also Fig.~\ref{fig:vscomp}); $v_{\mathrm{s,fit}}$: a value derived by fitting a pair of $v_{\mathrm{c}}$ and $v_{\mathrm{s}}$ in the approximative LOSVD given by Eq.~(\ref{eq:vlos}) (Sect.~\ref{sec:Compars}, point~\ref{item:app-LOSVD} and Fig.~\ref{fig:minfit}). The quantities with the superscript SS correspond to the subsample, where only measurements with two discernible inner peaks in the LOSVD are used.}
\label{tab:param-vs}
\end{table*}


\subsection{Recovering the potential from the simulated data}

We used a snapshot from our simulation, which 2.2\,Gyr after the decay of the cannibalized galaxy, as a source of the simulated data and tried to reconstruct the parameters of the potential of the host galaxy from the locations of the LOSVD peaks measured from the simulated data.

For a given host galaxy, the signal-to-noise (S$/$N) ratio in the simulated data is a function of the number of simulated particles, the age of the shell system, the distribution function of the cannibalized galaxy, and the impact velocity. For a given radius in the simulated data, we can obtain arbitrarily good or bad S$/$N ratios by tuning these parameters. Thus, we adopted the universal criteria: 1) the LOSVD of each shell is observed down to 0.9 times its radius; 2) we measured the positions of the LOSVD peaks in different locations within the shell, sampled by 1\,kpc steps. We do not estimate the errors, since the real data will be dominated by other sources. We quote only the mean square deviation and the standard error of the linear regression. 

\begin{figure*}
\centering
\includegraphics[width=8.5cm]{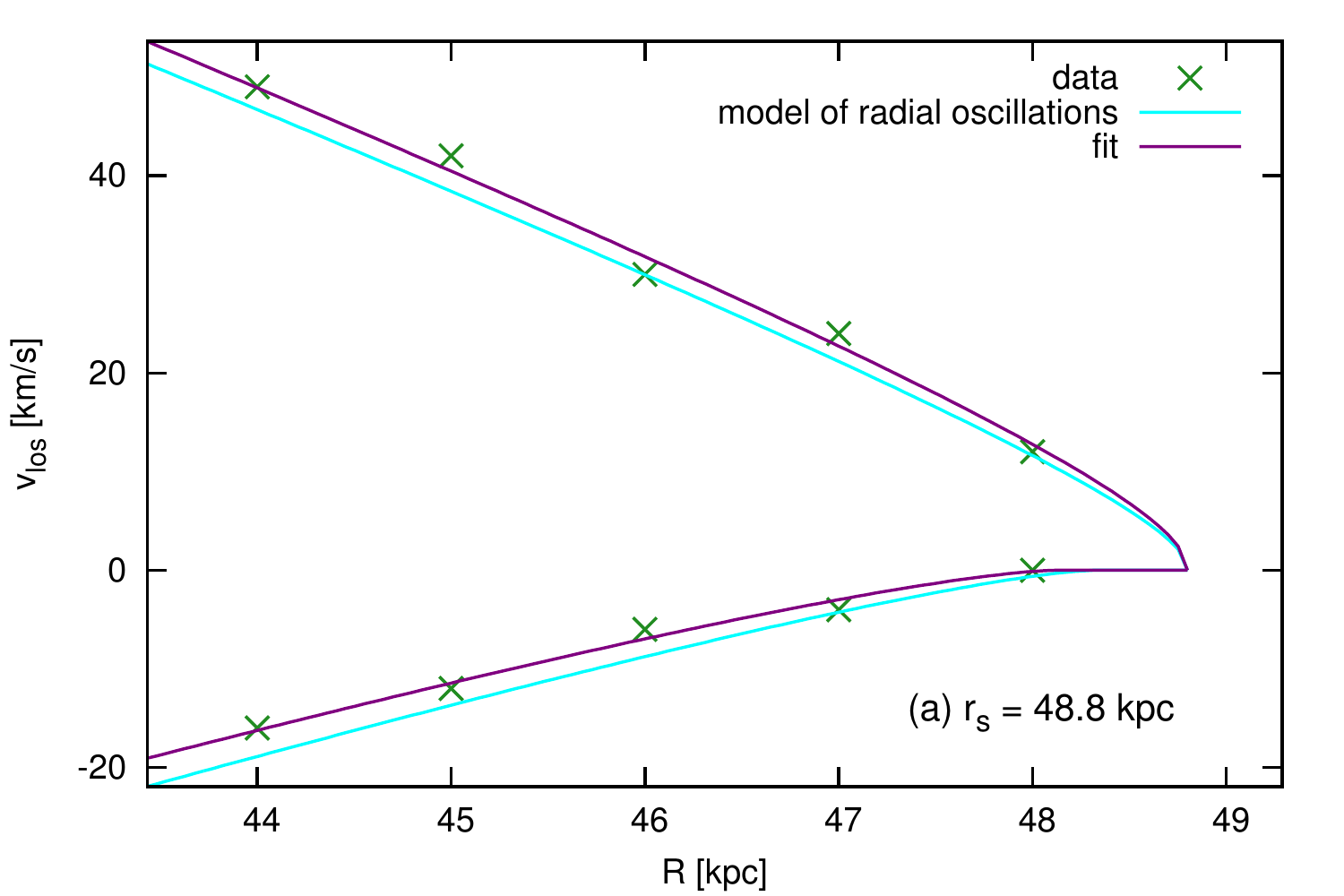}
\includegraphics[width=8.5cm]{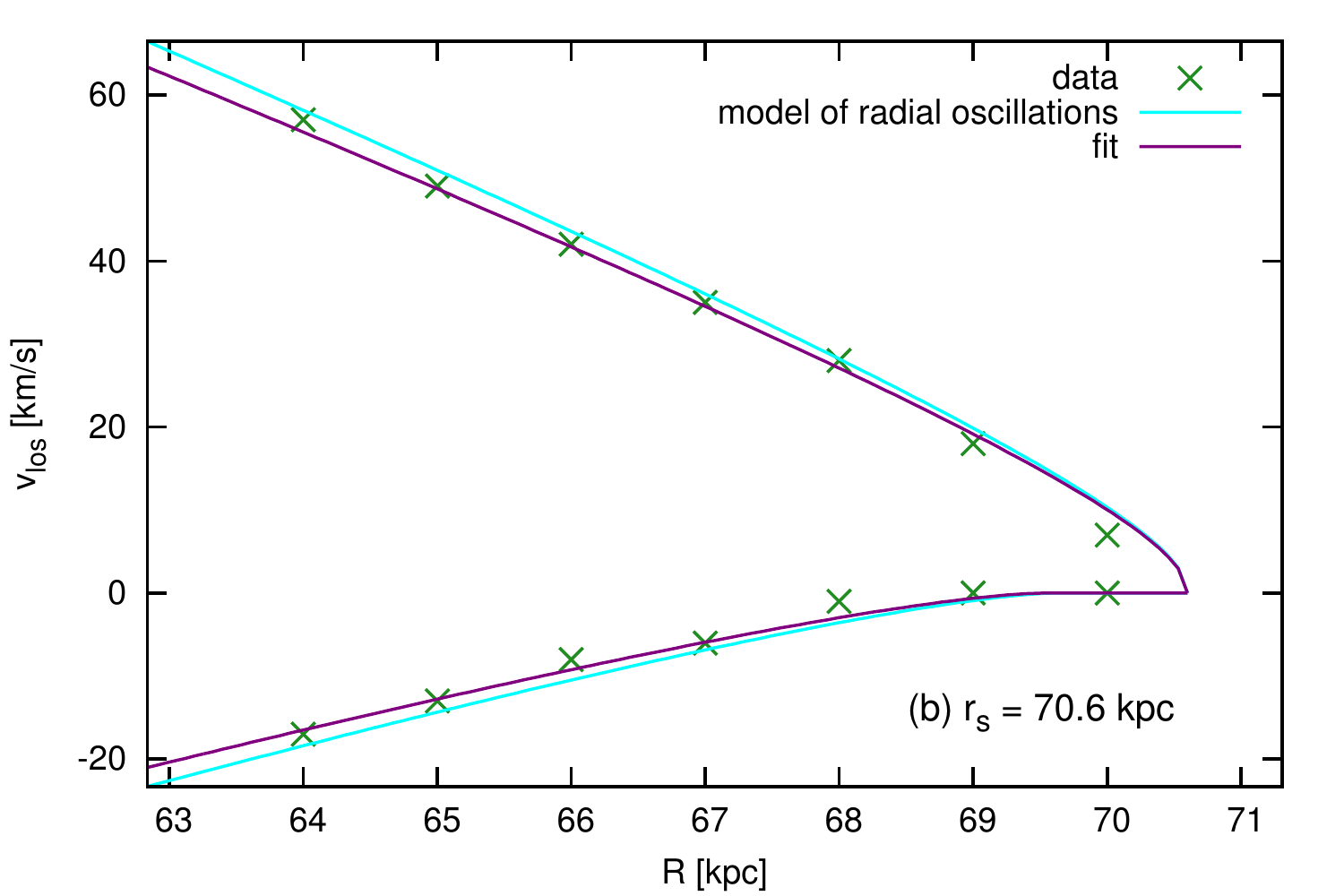}\\
\includegraphics[width=8.5cm]{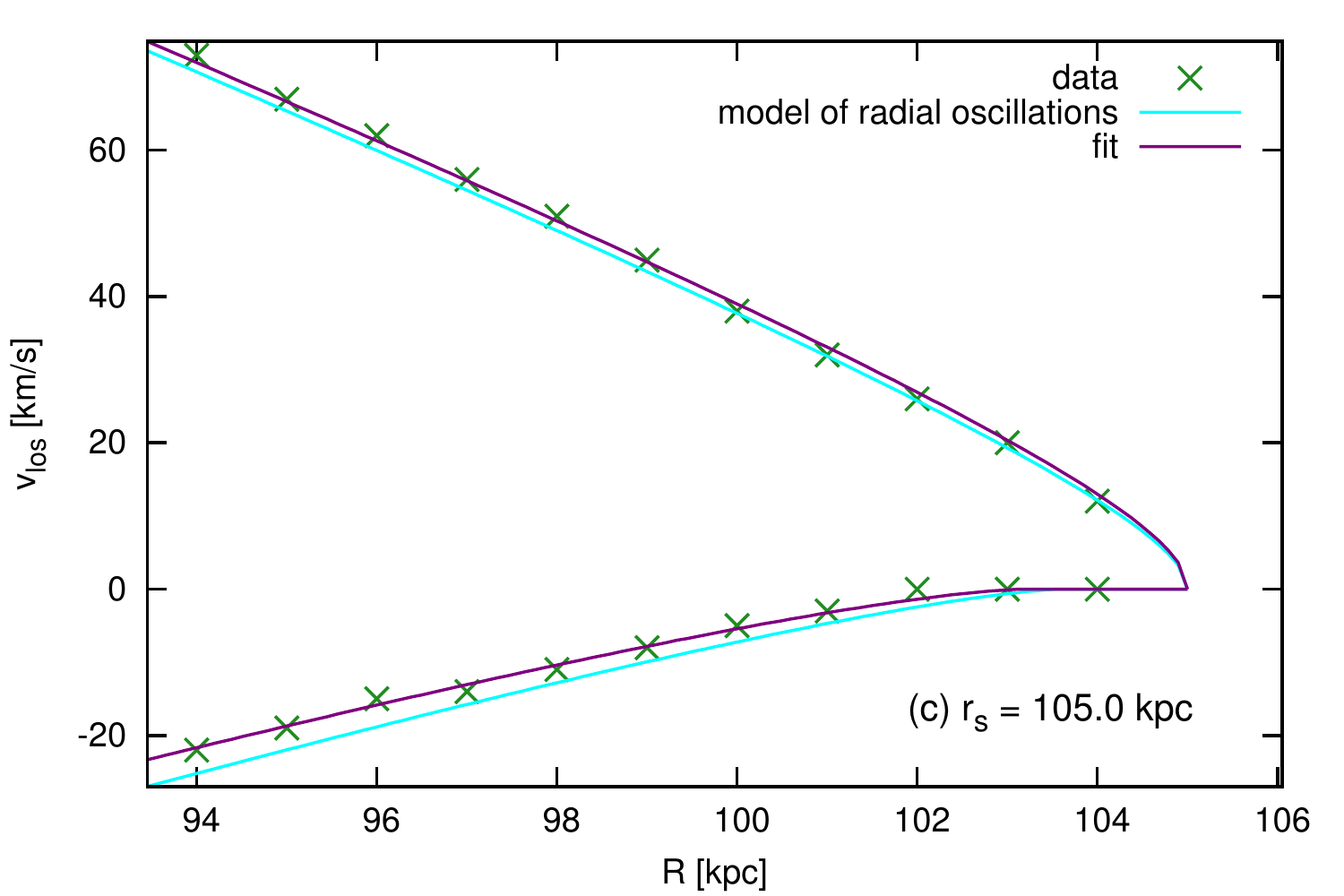}
\includegraphics[width=8.5cm]{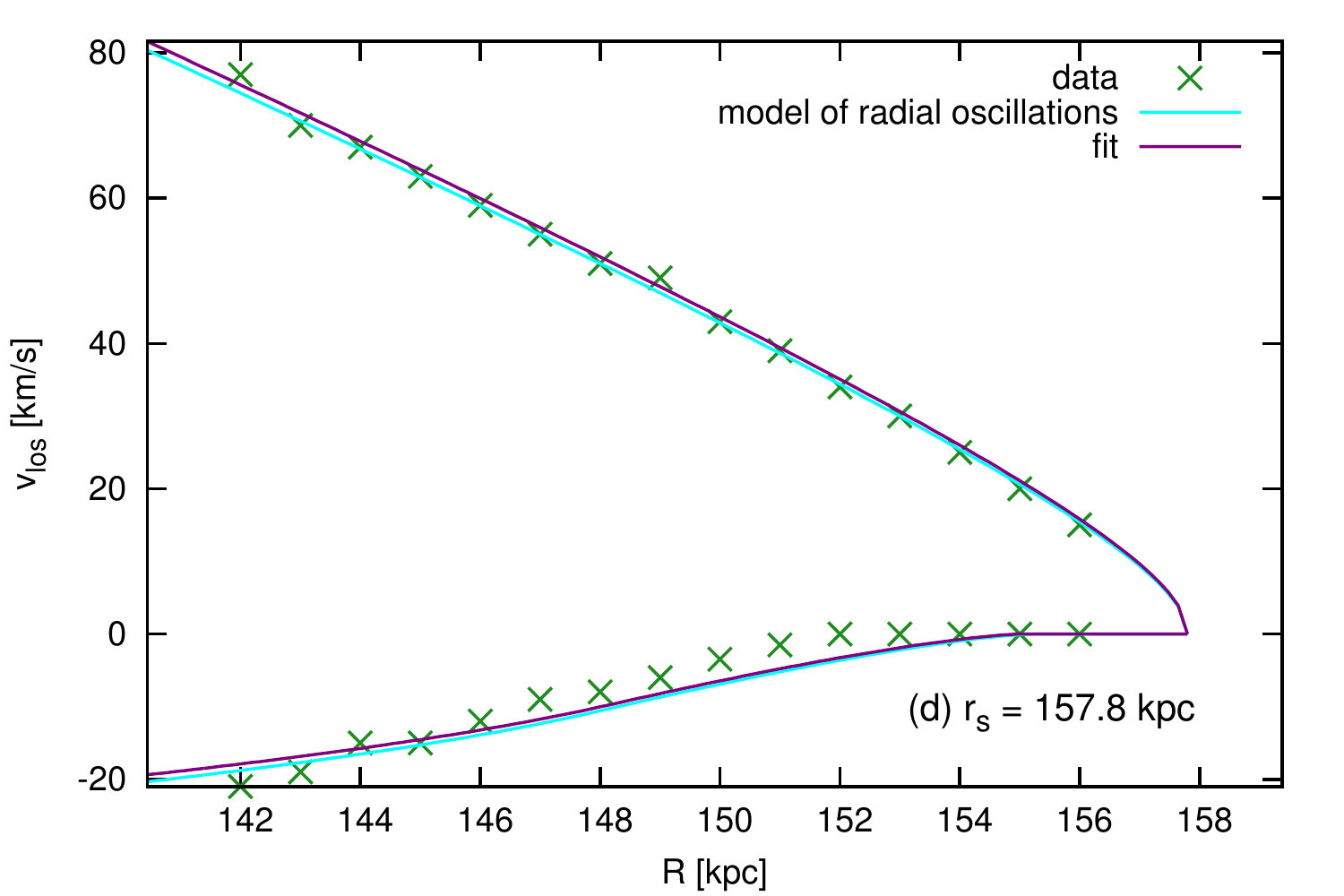}
\caption{Fits for circular velocity $v_{\mathrm{c}}$ and shell velocity $v_{\mathrm{s}}$ using the approximative LOSVD (Sect.~\ref{sec:Compars}, point~\ref{item:app-LOSVD}) given by Eq.~(\ref{eq:vlos}) for four shells ($r_{\mathrm{s}}$ indicated in bottom right corner of each plot) in the simulation 2.2\,Gyr after the decay of the cannibalized galaxy. The best fit is the purple curve, and its parameters are shown in Tables~\ref{tab:param-vc} and \ref{tab:param-vs} in the columns labeled $v_{\mathrm{c,fit}}$ and $v_{\mathrm{s,fit}}$. The green crosses mark the measured maxima in the LOSVD, and the light blue curves show the locations of the theoretical maxima derived from the host galaxy potential according to the model of radial oscillations (Sect.~\ref{sub:LOSVD-rad}). Note that the values of $v_{\mathrm{c}}$ and $v_{\mathrm{s}}$ used in the approximative LOSVD for the purple line were obtained by fitting the parameters to the simulated data, whereas in Figs.~\ref{fig:120app}, \ref{fig:app-rez}, and \ref{fig:rezy-sim}, the values are known from the model of the host galaxy potential.}
\label{fig:minfit}
\end{figure*}

These criteria give us between 7 and 15 measurements for a shell. Each measurement contains two values: the positions of the outer and inner peaks, $v_{\mathrm{los,max}+}$ and $v_{\mathrm{los,max}-}$, respectively, for each projected radius $R$ (see green crosses in Fig.~\ref{fig:minfit}).

First we used the approximative maximal LOS velocities given by Eqs.~(\ref{eq:vc,obs}) and (\ref{eq:vs,obs}) for a direct calculation of the circular velocity $v_{\mathrm{c,eq(\ref{eq:vc,obs})}}$ at the shell edge radius $r_{\mathrm{s}}$ and the current shell velocity $v_{\mathrm{s,eq(\ref{eq:vs,obs})}}$. These equations are the inverse of Eq.~(\ref{eq:vlos,max}), which corresponds to the model shown in orange lines in pictures throughout the text (Sect.~\ref{sec:Compars}, point~\ref{item:app-vmax}). Mean values from all the measurements for each shell are shown in Tables~\ref{tab:param-vc} and \ref{tab:param-vs}. 

\begin{figure}
\centering{}
\resizebox{\hsize}{!}{\includegraphics{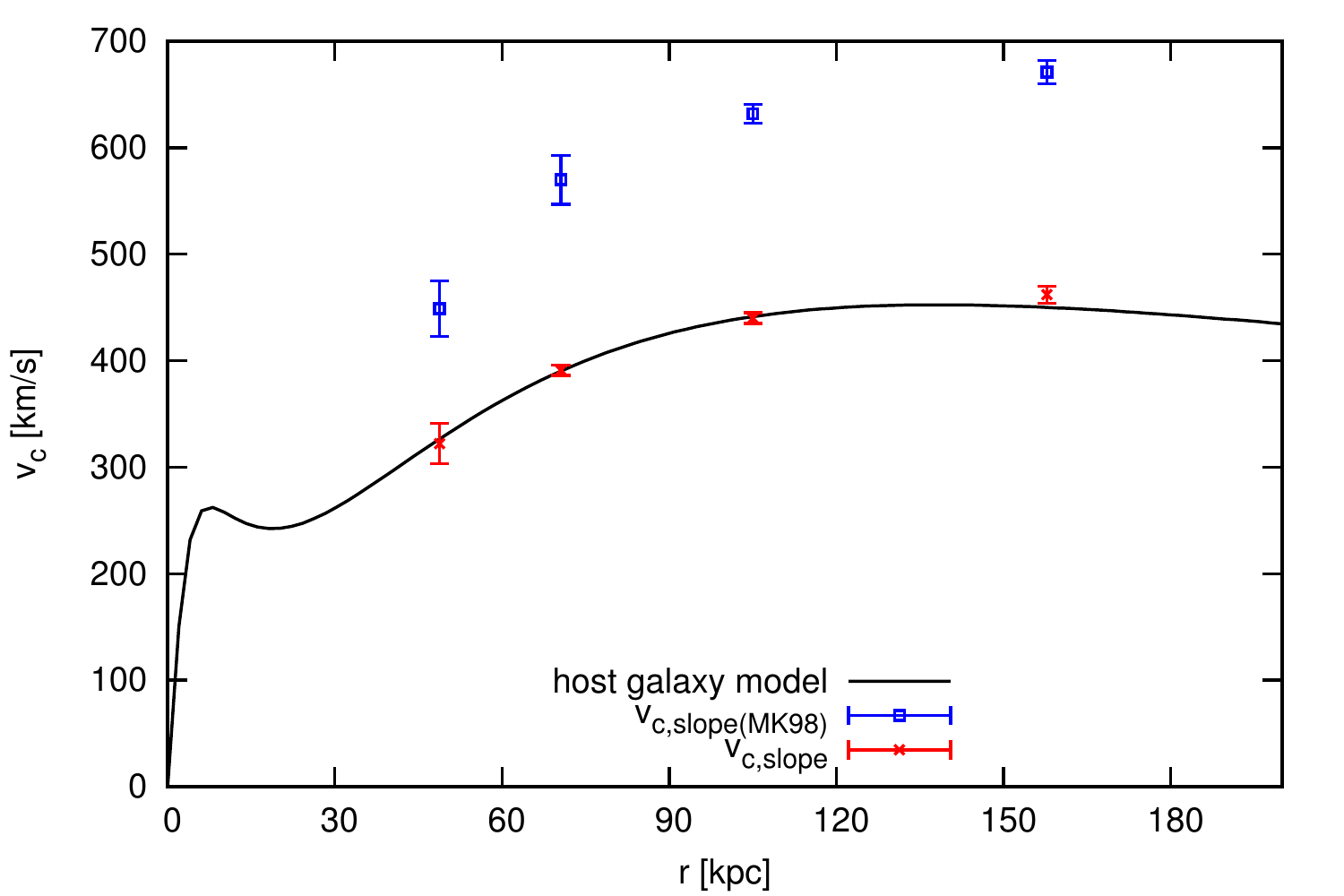}}
\caption{Circular velocity of the model and values derived from the simulated data: $v\mathrm{_{c,model}}$ of the host galaxy model is shown by the black line; blue and red points show values of circular velocity as they result from the analysis of the simulated LOSVD (see Sect.\,\ref{sec:sim-mod} and Table\,\ref{tab:param-vc} for the numbers).}
\label{fig:vccomp}
\end{figure}

We obtain a better agreement with the circular velocity of our host galaxy potential when using the slope of the LOSVD intensity maxima given by Eq.~(\ref{eq:sklon}), where we fit the linear function of the measured distance between the outer and the inner peak on the projected radius ($v_{\mathrm{c,slope}}$ in Table~\ref{tab:param-vc} and in Fig.~\ref{fig:vccomp}). 

\begin{figure}
\centering{}
\resizebox{\hsize}{!}{\includegraphics{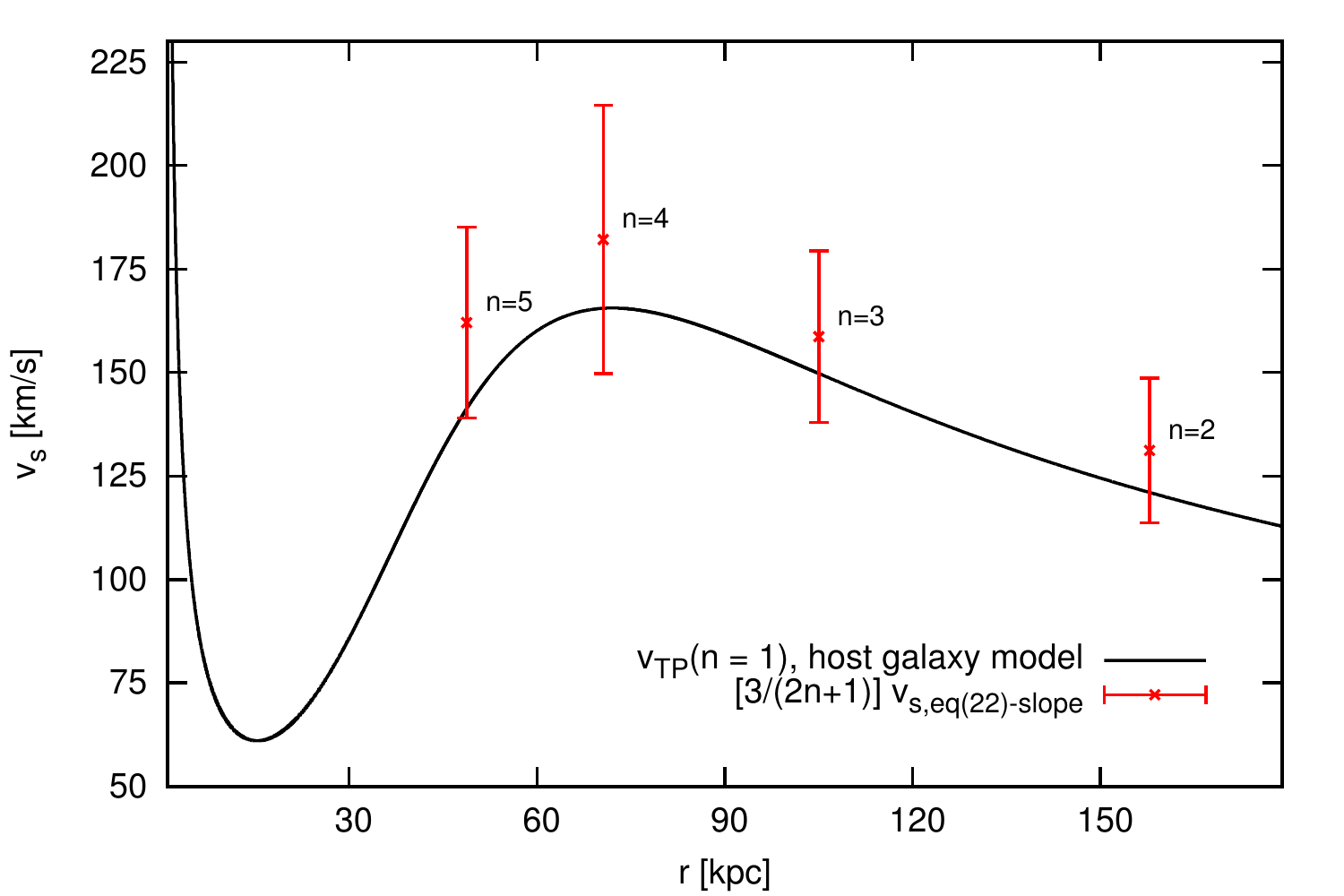}}
\caption{Comparison of velocity of the shell as a function of radius from the model and the simulated data. Velocity for the first shell ($n=1$) in the host galaxy model is shown by the black line. Red crosses show $v_{\mathrm{s,eq(\ref{eq:vs-vc})-slope}}$ (Table\,\ref{tab:param-vs}) as they result from the analysis of the simulated LOSVD. Values are corrected for shell number $n$ by the factor $3/(2n+1)$, so they correspond to velocity of the first shell (e.g., Eq.\,(\ref{eq:vTP})).}
\label{fig:vscomp}
\end{figure}

From the approximative maximal LOS velocities (Sect.~\ref{sec:Compars}, point~\ref{item:app-vmax}) given by Eq.~(\ref{eq:vlos,max}), we can derive a hybrid relation between the positions of the LOSVD peaks, the circular velocity at the shell edge radius $v_{\mathrm{c}}$, and the shell velocity:
\begin{equation}
v_{\mathrm{s}}^{2}=v_{\mathrm{c}}^{2}(1-R/r_{\mathrm{s0}})+\frac{v_{\mathrm{los,max}+}v_{\mathrm{los,max}-}}{4\left(R/r_{\mathrm{s0}}\right)^{2}\left(1+R/r_{\mathrm{s0}}\right)^{-2}-1}.
\label{eq:vs-vc}
\end{equation}
We substitute the values of $v_{\mathrm{c,slope}}$ derived from the measurements (that we know better describe the real circular velocity of host galaxy) into this relation, thus obtaining the improved measured shell velocity $v_{\mathrm{s,eq(\ref{eq:vs-vc})-slope}}$ (Table~\ref{tab:param-vs} and Fig.~\ref{fig:vscomp}).

In the zone between the current turning points and the shell edge, the inner peaks coalesce and gradually disappear (Fig.~\ref{fig:zona}). The simulated data do not show a disappearance of the inner peaks as abrupt and clear as the theoretical LOSVD profiles predict, so that in this zone, we can usually measure one inner peak at 0\,km$/$s. The information from these measurements is degenerate, and thus we defined a subsample of simulated measurements with all  four clear peaks in the LOSVD (in the columns labeled SS in Tables~\ref{tab:param-vc} and \ref{tab:param-vs}).

The spread of the values derived using the approximative maximal LOS velocities given by Eqs.~(\ref{eq:vc,obs}) and (\ref{eq:vs,obs}) is significantly lower for the subsample ($v_{\mathrm{c,eq(\ref{eq:vs,obs})}}^{\mathrm{SS}}$ and $v_{\mathrm{s,eq(\ref{eq:vc,obs})}}^{\mathrm{SS}}$) due to the exclusion of areas where these equations do not hold well. On the contrary, the slope of the linear regression in Eq.~(\ref{eq:sklon}) using the slope of the LOSVD intensity maxima gives a worse result (with a larger error) for the subsample $v_{\mathrm{c,slope}}^{\mathrm{SS}}$. 

The third option to derive the circular velocity $v_{\mathrm{c}}$ at the shell edge radius $r_{\mathrm{s}}$ and shell velocity $v_{\mathrm{s}}$ from the simulated data is to use the approximative LOSVD given by Eq.~(\ref{eq:vlos}), which corresponds to the model shown in purple lines in pictures throughout the text (Sect.~\ref{sec:Compars}, point~\ref{item:app-LOSVD}). However, this requires a numerical solution of the equation for a given pair of $v_{\mathrm{c}}$ and $v_{\mathrm{s}}$. We minimized the sum of sums of squared differences between $v_{\mathrm{los,max}\pm}(v_{\mathrm{c}},v_{\mathrm{s}})$ as given by the approximative LOSVD and the simulated data to obtain best fitted values $v_{\mathrm{c,fit}}$ and $v_{\mathrm{s,fit}}$ (see Tables~\ref{tab:param-vc} and \ref{tab:param-vs} for the results). Errors were estimated using the ordinary least squared minimization as if the functions $v_{\mathrm{los,max}+}(v_{\mathrm{c,fit}},v_{\mathrm{s,fit}})$ and $v_{\mathrm{los,max}-}(v_{\mathrm{c,fit}},v_{\mathrm{s,fit}})$ were fitted separately; quoted is the larger of the two errors.
The LOSVD intensity maxima resulting from this procedure are plotted in Fig.~\ref{fig:minfit}, together with the fitted data and the maxima given by the model of radial oscillations (Sect.~\ref{sub:LOSVD-rad}).

For the sake of comparison with the method of \citetalias{mk98}, we calculated the circular velocity $v_{\mathrm{c,slope(MK98)}}$ at the shell edge radius $r_{\mathrm{s}}$ using the slope of the LOSVD intensity maxima given by Eq.~(\ref{eq:sklon}). To mimic the measurement of the circular velocity according to the Eq.~(7) in \citetalias{mk98}, which was derived for the double-peaked profile, we assume $\bigtriangleup v_{\mathrm{los}}$ is the distance between the two outer peaks of the LOSVD intensity maxima. In Table~\ref{tab:param-vc} and Fig.~\ref{fig:vccomp}, we can easily see that the values $v_{\mathrm{c,slope(MK98)}}$ differ from the actual circular velocity of the host galaxy $v\mathrm{_{c,model}}$ by a factor of 1.3--1.5.

\subsection{Instrumental LOSVD}\label{sec:fwhm}

\begin{figure*}
\centering
\resizebox{\hsize}{!}{\includegraphics{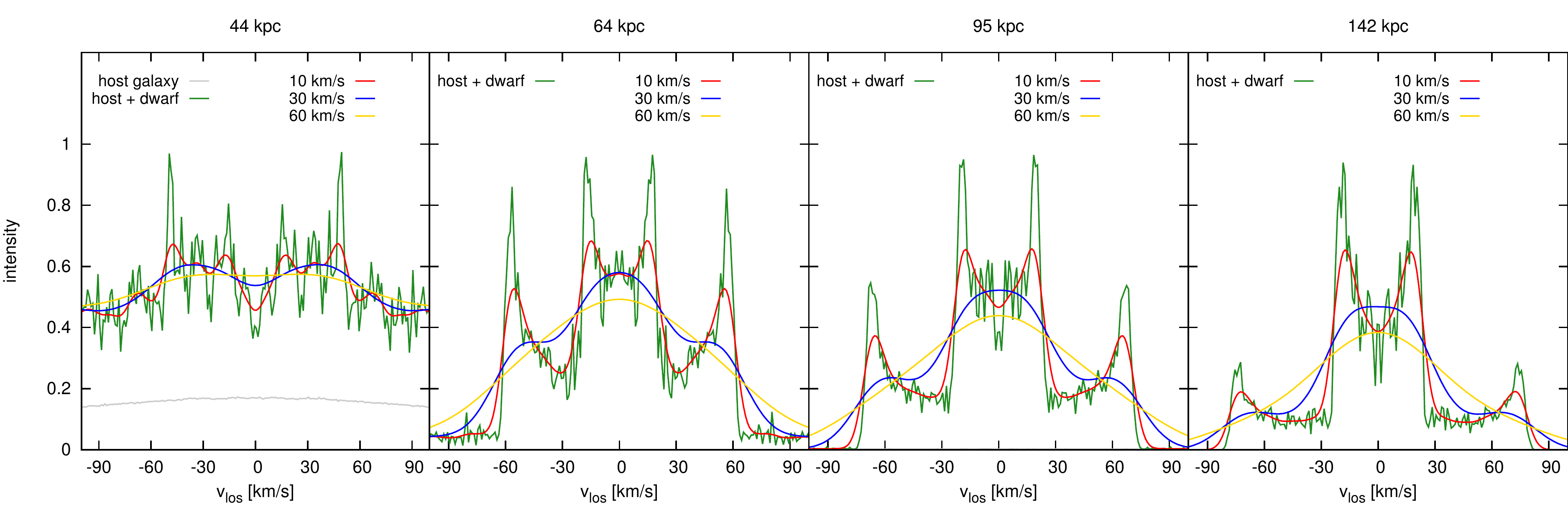}}
\caption{Line profiles of four shells at projected radii $0.9r_{\mathrm{s}}$ (indicated as the title of each plot, same as in Fig.~\ref{fig:rezy-sim}) 2.2\,Gyr after the decay of the cannibalized galaxy: gray lines show the LOSVDs for the host galaxy at a given radius (except for the radius of 44\,kpc the signal of the host galaxy is negligible comparing to the signal from the cannibalized galaxy); green lines show the total LOSVDs from the host and the cannibalized galaxy together; red, blue, and yellow lines show convolutions of the total simulated data with different Gaussians representing the instrumental profiles having the FWHM 10, 30, and 60\,km$/$s, respectively. Scaling is relative, similar as in Fig.~\ref{fig:rezy-sim}.}
\label{fig:konv}
\end{figure*}

When observed, the LOSVD is always influenced by instrumental dispersion, which naturally smoothes features of the spectral profile. In Fig.\,\ref{fig:konv}, we show the LOSVDs from the simulated data smoothed with different Gaussians representing the instrumental profiles having the full width at half maximum (FWHM) of 10, 30, and 60\,km$/$s. It is obvious that relatively high spectral resolution is necessary for observing an imprint of shell peaks in line profiles.


\section{Discussion}

We developed a new method to measure the potential of shell galaxies from kinematical data, extending the work of \citetalias{mk98}. The method splits into three different analytical and semi-analytical approaches for obtaining the circular velocity in the host galaxy, $v_{\mathrm{c}}$, and the current shell phase velocity, $v_{\mathrm{s}}$: 

\begin{enumerate}
\item The approximative LOSVD: using Eq.~(\ref{eq:vlos}) and an assumption of the behavior of the shell brightness as a function of the shell radius (Sect.~\ref{sub:LOSVD-app}).
\label{item:dis-LOSVD}

\item The approximative maximal LOS velocities: Eqs.~(\ref{eq:vc,obs}) and (\ref{eq:vs,obs}) (Sect.~\ref{sub:vlos,max}).
\label{item:dis-los}

\item Using the slope of the LOSVD intensity maxima in the \mbox{$R \times v_{\mathrm{los}}$} diagram in Eq.~(\ref{eq:sklon}) (Sect.~\ref{sub:vlos,max}).
\end{enumerate}

In Sect.~\ref{sec:Compars}, the first two approaches are compared to the model of radial oscillations (numerical integration of radial trajectories of stars in the host galaxy potential, Sect.~\ref{sub:rad_osc}). All three approaches are then applied to data for the four shells obtained from a test-particle simulation and compared to the theoretical values (Sect.~\ref{sec:sim-mod}). 

Approach~\ref{item:dis-LOSVD} requires a numerical solution to Eq.~(\ref{eq:vlos}) and the search for a pair of $v_{\mathrm{c}}$ and $v_{\mathrm{s}}$, which matches the (simulated) data best. Although this approach is not limited by any assumptions about the radius of the maximal LOS velocity (Sect.~\ref{sub:rvmax}), it does not give a better estimate of $v_{\mathrm{c}}$ and $v_{\mathrm{s}}$ for our simulated shell galaxy than the other two methods. The deviation from the real value of $v_{\mathrm{c}}$ is between 2\,\% and 6\,\%.

Using the approximative maximal LOS velocities approach~\ref{item:dis-los}, results in simple analytical relations and is the only one that can in principle be used for an LOSVD measured at only one projected radius. Nevertheless, when measuring in the zone between the radius of the current turning points and the shell radius, we can expect very bad estimates of $v_{\mathrm{c}}$ and $v_{\mathrm{s}}$. The mean value from more measurements of the LOSVD peaks for each shell of our simulated shell galaxy has similar accuracy to those of approach~\ref{item:dis-LOSVD}, provided that we include only the measurements outside the zone between the radius of the current turning points and the shell radius.

The best method for deriving the circular velocity in the potential of the host galaxy seems to be to use the slope of the LOSVD intensity maxima, with a typical deviation in the order of units of km$/$s when fitting a linear function over all the measured positions of the LOSVD peaks for each shell. This circular velocity is then used in the hybrid relation, Eq.~(\ref{eq:vs-vc}), to obtain the best estimate of the shell velocity. 

All the approaches, however, derive a shell velocity systematically larger than the prediction of the model of radial oscillations $v_{\mathrm{s,model}}$ and the value derived from positions between the times 2.49--2.51\,Gyr in the simulation $v_{\mathrm{s,sim}}$ (Table~\ref{tab:param-vc}). This is because the simulated LOSVD peaks lie too far out (for the outer peaks) or too far in (for the inner peaks) when compared to the model of radial oscillations. That can be caused by nonradial trajectories of the stars of the cannibalized galaxy or by poor definition of the shell radius in the simulation. 

Nevertheless, the shell velocity depends, even in the simplified model of an instant decay of the cannibalized galaxy in a spherically symmetric host galaxy (Sec.~\ref{sub:rad_osc}), on the serial number of the shell $n$ and on the whole potential from the center of the galaxy up to the shell radius (see Eq.~(\ref{eq:vTP})). A comparison of its measured velocity to theoretical predictions is possible only for a given model of the potential of the host galaxy and the presumed serial number of the observed shells. In such a case, however, it can be used to exclude some parameters or models of the potential that would otherwise fit the observed circular velocity.

The first shell has a serial number equal to one. A higher serial number means a younger shell. On the same radius, the velocity of each shell is always smaller than that of the previous one. In practice, it is difficult to establish whether the outermost observed shell was the first one created, or whether the first shell (or even the first couple of shells) were already unobservable. Here, we can use the potential derived from our method or a completely different one in a reverse way: to determine the velocity of the first shell on the given radius and to compare it to the velocity derived from the positions of the LOSVD peaks. Knowing the serial number of the outermost shell and its position allows us then to determine the time from the merger and the impact direction of the cannibalized galaxy. Moreover, the measurement of shell velocities reveals the presence of shells from different generations \citep{katka11}.

Our method for measuring the potential of shell galaxies has several limitations. Theoretical analyses were conducted over spherically symmetric shells, while the test-particle simulation was run for a strictly radial merger and analyzed in a projection plane parallel to the axis of the merger. In addition, both analytical analysis and simulations assume spherical symmetry of the potential of the host galaxy. In reality, the regular shell systems with more shells in one galaxy are more often connected to galaxies with significant ellipticity \citep{dupraz86}. Moreover, in cosmological simulations with cold dark matter, halos of galaxies are described as triaxial ellipsoids \citep[e.g.,][]{jing02,bailin05,allgood06}. However, the effect of the ellipticity of the isophotes of the host galaxy on the shell kinematics need not be dramatic, as the shells have the tendency to follow equipotentials that are in general less elliptical than the isophotes. \citet{dupraz86} concluded that while the ellipticity of observed shells is generally low, it is neatly correlated to the eccentricity of the host galaxy. \citet{prieur88} pointed out that the shells in NGC\,3923 are much rounder than the underlying galaxy and have an ellipticity that is similar to the inferred equipotential surfaces. This idea was originally put forward by \citet{dupraz86}, who found such a relationship for their merger simulations. Our method is in principle applicable even to shells spread around the galactic center, which are usually connected to rounder elliptical galaxies if they were created in a close-to-radial merger. Nevertheless, the combination of the effects of the projection plane, merger axis, and ellipticity of the host galaxy can modify our results and require further analyses. 

Because the kinematics of the stars that left the cannibalized galaxy is in the first approximation a test-particle problem, they should not be much affected by self-gravity of the cannibalized galaxy and the dynamical friction that this galaxy undergoes during the merger, both of which have been neglected in this work.

Another complication is that the spectral resolution required to distinguish all four peaks is probably quite high (Sect.~\ref{sec:fwhm} and Fig.~\ref{fig:konv}) and the shell contrast is usually small. Nevertheless, there is the possibility to measure shell kinematics using the LOS velocities of individual globular clusters, planetary nebulae, and, in the Local group of galaxies, even of individual stars.


\section{Conclusions}

Kinematics of regular shells produced during nearly radial minor mergers of galaxies can be used to constrain their gravitational force field and thus the dark matter distribution. \citet{mk98} showed that the LOSVD measured near the edges of a shell has a double-peaked shape, and found a relation between the values of the two LOS velocity peaks and the circular velocity. Their approximation is limited to stationary shells.

We have extended their theoretical analysis to traveling shells. We find that in two-component giant galaxies with realistically massive dark matter halos, shell propagation velocity is significantly higher, typically 30--150 km$/$s, compared to values quoted in the theoretical studies in the literature. We show that such large speeds have considerable impact on the LOS kinematics of shells. We demonstrate that each peak of the double-peaked profile is split into two, producing a quadruple-peaked LOSVD. We derive a new approximation, relating the circular velocity of the host galaxy potential at the shell edge radius, as well as the current phase velocity of the shell, to the positions of the four peaks.

In galaxies with multiple shells, we can use circular velocities measured by these methods to determine the potential of the host galaxy over a large span in radii, whereas the measured shell phase velocity carries information on the age of the shell system, and the arrival direction of the cannibalized galaxy. The potential observation of multigeneration shell systems contains additional limits on the shape of the potential of the host galaxy.


\begin{acknowledgements}
We acknowledge support from the following sources: grant No.\,205/08/H005 by Czech Science Foundation (IE, LJ, MK, KB, and TS); grant MUNI/A/0968/2009 by Masaryk University in Brno (LJ, KB, and TS); research plan AV0Z10030501 by Academy of Sciences of the Czech Republic (IE, BJ, MK, MB, KB, and IS); grant LC06014\,--\,Center for Theoretical Astrophysics\,--\,by Czech Ministry of Education (BJ and IS); and the project SVV261301 by Charles University in Prague (IE, MK, and MB). This work has been done with the support for a long-term development of the research institution RVO67985815 (IE, BJ, MB, and KB). LJ acknowledges the support by the 2-years ESO PhD studentship, held in ESO, Santiago.
\end{acknowledgements}

\bibliographystyle{aa}
\bibliography{shells}


\end{document}